\documentclass[journal]{IEEEtran}

\usepackage{ucs}
\usepackage[utf8x]{inputenc}
\usepackage[cmex10]{amsmath}
\usepackage{cite, amsfonts, amssymb, amsthm, bm, bbm, graphicx, relsize, multirow, booktabs, tikz,soul}
\usepackage[american]{babel}
\usepackage[T1]{fontenc}
\usepackage{algorithmic, algorithm}
\usepackage[multiple]{footmisc}
\usepackage{ucs}
\usepackage[utf8x]{inputenc}
\usepackage[cmex10]{amsmath}
\usepackage[american]{babel}
\usepackage[T1]{fontenc}
\usepackage{algorithmic, algorithm}
\usepackage[multiple]{footmisc}
\usepackage{tabulary}
\usepackage{subcaption}
\usepackage[font={small}]{caption}
\usepackage{float}
\theoremstyle{definition}

\theoremstyle{remark}

\newcommand{\ds}{\displaystyle}

\begin{document}
% comando per disabilitare i dashes quando i nomi degli autori si ripetono (IEEEexample:BSTcontrol definito in finalRefs.bib)
\bstctlcite{IEEEexample:BSTcontrol}

This paper was submitted for publication on the \textit{IEEE Transactions on Communications} on November 3, 2018  and was assigned reference number TCOM-TPS-18-1223.
It was finally accepted for publication on June 20, 2019.  

\bigskip

\copyright 2019 IEEE. Personal use of this material is permitted. Permission from IEEE must be obtained for all other uses, in any current or future media, including reprinting/republishing this material for advertising or promotional purposes, creating new collective works, for resale or redistribution to servers or lists, or reuse of any copyrighted  component of this work in other works.”

\title{Subspace Tracking and Least Squares Approaches to Channel Estimation in Millimeter Wave Multiuser MIMO}
\author{Stefano Buzzi, {\em Senior Member}, {\em IEEE},  and Carmen D'Andrea, {\em Student Member}, {\em IEEE}
\thanks{This paper has been partly presented at the 
21th International ITG Workshop on Smart Antennas, Berlin (Germany), March 2017.
}\thanks{The authors are with the Department of Electrical and Information Engineering, University of Cassino and Lazio Meridionale, I-03043 Cassino, Italy (buzzi@unicas.it, carmen.dandrea@unicas.it).}
}
\maketitle
\begin{abstract}
The problem of MIMO channel estimation at  millimeter wave frequencies, both in a single-user and in a multi-user setting, is tackled in this paper. Using a subspace approach, we develop a protocol enabling the estimation of the right (resp. left) singular vectors at the transmitter (resp. receiver) side; then, we adapt the projection approximation subspace tracking with deflation  and the orthogonal Oja algorithms to our framework and obtain two channel estimation algorithms. We also present an alternative algorithm based on the least squares  approach. The hybrid analog/digital nature of the beamformer is also explicitly taken into account at the algorithm design stage. 
In order to limit the system complexity, a fixed analog beamformer is used at both sides of the communication links.
The obtained numerical results, showing the accuracy in the estimation of the channel matrix dominant singular vectors, the system achievable spectral efficiency, and the system bit-error-rate,  prove that the proposed algorithms are effective, and that they compare favorably, in terms of the performance-complexity trade-off, with respect to several competing alternatives. 
\end{abstract}


\section{Introduction}

The use of frequency bands in the range 10−100 GHz, a.k.a. millimeter waves (mmWaves), for cellular communications, is one of the main technological innovations brought by fifth generation (5G) wireless networks \cite{whatwillbe}. 
Indeed, the scarcity of available frequency bands in the sub-6 GHz spectrum has been a strong thrust for considering the use of higher frequencies for cellular applications. A recent research \cite{itwillwork} has shown that mmWaves, despite increased path-loss and atmospheric absorption phenomena, can be actually used for cellular communications over short-range distances (up to 100-200 meters), provided that multiple antennas are used at both sides of the communication link. It thus follows that multiple-input multiple-output (MIMO) processing is one distinguishing and key feature of mmWave systems.
MIMO channels at mmWave behave considerably different from MIMO channels at conventional sub-6 GHz cellular frequencies \cite{ztepaper,WCM_paper}. Indeed, while at the latter frequencies the rich scattering environment leads to channel matrices that result from the superposition of path-loss effects, shadowing, and small-scale fading independently distributed on each antenna element, at mmWave propagation happens through possible line-of-sight and/or through one-hop reflected paths, and blockage effects are more frequent. Given these differences, conventional MIMO channel estimation algorithms developed for sub-6 GHz cellular frequencies are not suited to perform channel estimation at mmWave, which instead can leverage the sparse and parametric structure of the channel. On top of this, complexity of the transceiver hardware at mmWave has led to the adoption of hybrid (HY) analog/digital beamforming structures \cite{heath2016overview}, wherein the number of RF transceiver chains is smaller than the number of antenna elements, and this of course poses some constraints on the signal processing algorithms that can be used in the channel estimation algorithms \cite{mendez2015channel}. The need to use the large antenna arrays at mmWave frequencies makes the task of channel estimation even more challenging, since this leads to an increase of the size of the channel matrix to be estimated. 

Channel estimation and/or principal directions estimation for MIMO mmWave channels has been a very active research area in the last few years. 
The paper \cite{alkhateeb2014channel} was one of the first to propose a channel estimation scheme taking into account the sparse nature of the MIMO mmWave channel and the presence of HY beamformers. In \cite{ghauch2016subspace} the authors exploit the time division duplex (TDD) protocol and adopt a subspace approach, based on the Arnoldi iteration, to propose a new channel estimation scheme. The papers \cite{haghighatshoar2016low,haghighatshoar2016massive} propose several algorithms for estimating signal subspaces based on low-dimensional projections, a situation encountered in mmWave channels wherein the use of analog combining schemes leads to reduced-dimensionality observations. 
Channel estimation for mmWave MIMO channels is also considered in \cite{kokshoorn2017millimeter}, wherein a fast channel estimation  algorithm based on an overlapped beam pattern design procedure is proposed, and in \cite{ma2017channel}, wherein a procedure for 3-D lens antenna arrays is developed. The papers \cite{mo2014channel, mo2017channel,mo2017hybrid} focus instead on the problem of channel estimation for the case in which low-resolution analog-to-digital converters are used in place of the analog combining stage. The sparse nature of the MIMO channel at mmWave is exploited in papers \cite{Choi_Beam2015,Song_TWC2018,Song_TWC2019}, that resort to compressed sensing techniques to perform beam alignment; while \cite{Choi_Beam2015} adopts the usual compressed sensing approach based on the minimization of the $\ell_1$-norm, papers \cite{Song_TWC2018, Song_TWC2019} instead use the recently developed non-negative least squares algorithm, which, under suitable conditions, implicitly promotes the sparsity of the sought solution. In order to circumvent the problems that arise due to the usual channel angles quantization, paper 
\cite{Hu_TVT2018} proposes an iterative super-resolution channel estimation scheme that, using the gradient descent method, moves the angles estimates from the initial on-grid value to  better off-grid values. Finally, the paper 
\cite{WU_CL2018} considers the problem of mmWave MIMO channel estimation in the presence of transmitter-receiver hardware impairments, that are modeled as an additional Gaussian-distributed vector-valued disturbance with power proportional to the transmit power at each antenna. 

Following on this research track, and building upon the conference paper 
\cite{buzzi_WSA2017}, this paper considers the problem of estimating the principal left and right singular vectors of a mmWave MIMO channel wherein HY analog/digital beamforming is used. The proposed algorithms are presented both for the single-user and the multi-user scenario, wherein a base station (BS) simultaneously learns the channel from several mobile stations (MSs). The contribution of this paper can be summarized as follows.

\begin{enumerate}
\item
Using a subspace approach, we develop a protocol enabling the estimation of the right (resp. left) singular vectors at the transmitter (resp. receiver) side; then, we adapt the \textit{Projection Approximation Subspace Tracking with deflation} (PASTd) algorithm  \cite{yang1995projection}, and the \textit{Orthogonal Oja} (OOJA) algorithm \cite{abed2000orthogonal} to our framework and obtain two subspace-based channel estimation algorithms.
\item Using the parametric structure assumption for the mmWave channel matrix, we develop an LS approach to the channel estimation.
\item
We adapt the proposed algorithms in order to take into account the HY analog/digital beamforming structure usually employed in mmWave wireless links. In particular, we assume that both at the transmitter and at the receiver the front-end RF modules are made by an analog  combining matrix whose columns correspond to beam-steering array response vectors, and show that the proposed channel estimation algorithms based on subspace tracking can be applied in this scenario too. 
\item
We compare the proposed algorithms with other competing alternatives available in the open literature, namely the   \textit{Approximate Maximum Likelihood} (AML) algorithm in \cite{haghighatshoar2016low,haghighatshoar2016massive}, the \textit{Subspace Estimation using Arnoldi iteration} (ARN) in \cite{ghauch2016subspace} and the \textit{Adaptive Estimation (AE)} algorithm in \cite{alkhateeb2014channel}. In this context, we also generalize the AML algorithm -- presented in \cite{haghighatshoar2016low,haghighatshoar2016massive} for the case in which the mobile station (MS) has just one antenna -- to the case in which the MS is equipped with multiple antennas.

\item We also focus on the problem of joint multiuser MIMO channel estimation using mmWave frequencies and the TDD protocol. We develop a framework wherein first the BS sends a suitable probing signal in order to let the MSs estimate the dominant left channel eigenvectors; then, the MSs, using the estimated vectors as pre-coding beamformers, send pilot sequences to enable channel estimation at the BS.

\item
We also show how the proposed channel estimation schemes can be used to implement a simple pilot-less differential modulation scheme and the corresponding symbol error probability (SER) is numerically evaluated. 
\end{enumerate}

This paper is organized as follows. Next section contains the description of the channel model in the single user and multiuser scenarios; in Section III we explain the subspace-based channel estimation in the single user scenario, using both the approaches of fully digital (FD) and HY beamforming, and introducing the PASTd and OOJA algorithms. In Section IV we describe the LS channel estimation scheme for a single-user scenario, while Section V is devoted to the generalization of the proposed algorithms to the multiuser case. Section VI  is devoted to the definition of the performance measures used in the paper and to the discussion of the numerical results. Finally, concluding remarks are given in Section VII.

\textit{Notation:} The following notation is used in the paper. Bold lowercase letters (such as $\mathbf{a}$) denote column vectors, bold uppercase letters (such as $\mathbf{A}$) denote matrices, non-bold letters $a$ and $A$ denote scalar values. The transpose, the inverse and the conjugate transpose of a matrix $\mathbf{A}$ are denoted by $\mathbf{A}^T$, $\mathbf{A}^{-1}$ and $\mathbf{A}^H$, respectively. The trace of the matrix $\mathbf{A}$ is denoted as tr$\left(\mathbf{A}\right)$. The $i$-th entry of the vector $\mathbf{a}$ is denoted as $\left[\mathbf{a}\right]_i$ and the $(i,j)$-th entry of the matrix $\mathbf{A}$ is $\left[\mathbf{A}\right]_{i,j}$. The real part operator is denoted as $\mathfrak{R}\{ \cdot \}$. The $N$-dimensional identity matrix is denoted as $\mathbf{I}_N$ and the $N$-dimensional diagonal matrix with elements $\alpha_1, \ldots, \alpha_N$ is denoted as diag$(\alpha_1, \ldots, \alpha_N)$. The statistical expectation operator is denoted as $\mathbb{E}[\cdot]$; $\mathcal{CN}\left(\mu,\sigma^2\right)$ denotes a complex circularly symmetric Gaussian random variable with mean $\mu$ and variance $\sigma^2$, while $\mathcal{U}(a,b)$ denotes a random variable that is uniformly distributed in $[a,b]$.

\begin{center}
\begin{figure*}[t]
\includegraphics[scale=0.26]{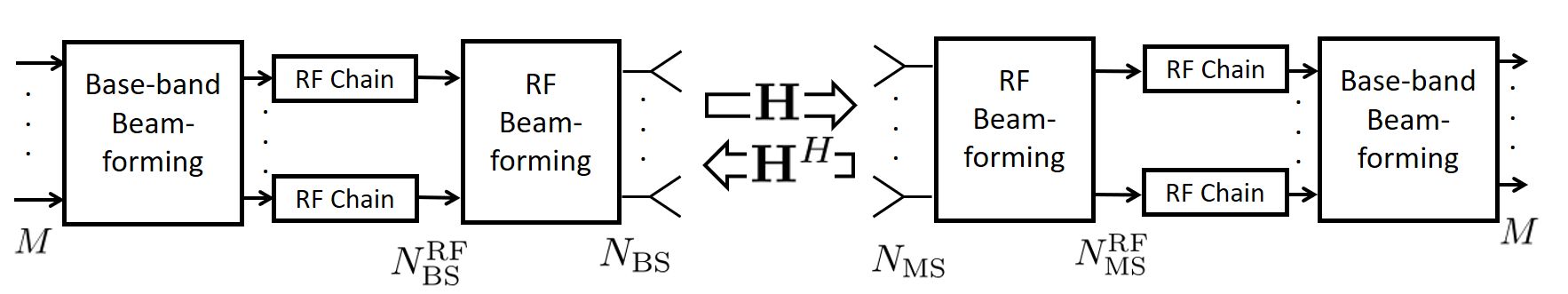}
\caption{Block scheme of the considered transceiver architecture.}
\label{fig:scheme}
\end{figure*}
\end{center}

\section{The system model}
We consider a single-cell multiuser mmWave wireless system with one BS and $K$ MSs. 
The time-division-duplex (TDD) protocol is implemented, so that the BS-to-MS channel is the conjugate transpose of the reciprocal MS-to-BS channel, provided that the transmission time does not exceed the channel coherence interval.
We denote by $N_{\rm BS}$ the number of antennas at the BS, and by $N_{\rm MS}$ the number of antennas at the MSs.
Although all of the techniques that will be presented in the paper can be applied to transceivers using  any antenna array configuration, for the sake of simplicity, a 
bi-dimensional model is assumed, and both the BS and the MS are assumed to be equipped with a uniform linear array (ULA). 
The number of parallel data streams, i.e. the multiplexing order, is denoted by $M$, and is assumed to be the same for all the MSs; in particular, the BS will send in parallel $MK$ data streams to the MSs in the same time-frequency slot, while, on the uplink, each MS will send $M$ parallel data streams.  
A schematic representation of the considered system, for a single MS-BS link,  is reported in Figure \ref{fig:scheme}. In particular, beamforming at both sides of the link is of the HY type, in the sense that, in order to reduce hardware complexity, analog signal combining is performed at RF in the transceiver front-end in order to reduce the dimensionality of the signals, and, then FD baseband combining is performed. We will denote by 
$N_{\rm BS}^{\rm RF}<N_{\rm BS}$ and $N_{\rm MS}^{\rm RF}<N_{\rm MS}$  the number of RF chains at the BS and at each MS, respectively.

We denote by $\mathbf{H}_k$ the $(N_{\rm MS} \times N_{\rm BS})$-dimensional matrix representing the channel from the BS to the $k$-th MS. Due to  TDD operation the reverse-link propagation channel is expressed as $\mathbf{H}_k^H$.
According to the popular narrowband clustered mmWave channel model 
\cite{spatiallysparse_heath,cairewsa2016,lee2014exploiting}, the channel  matrix $\mathbf{H}_k$ is expressed as:
\begin{equation}
\begin{array}{lll}
\mathbf{H}_k &= \gamma \ds \sum_{i=1}^{N_{\rm cl}}\ds \sum_{l=1}^{N_{{\rm ray},i}}\alpha_{i,l,k}
\sqrt{L(r_{i,l,k})} \mathbf{a}_{\rm MS}(\phi_{i,l,k}^{\rm MS}) \mathbf{a}_{\rm BS}^H(\phi_{i,l,k}^{\rm BS}) \\ & + \mathbf{H}_{k, {\rm LOS}}\; .
\end{array}
\label{eq:channel1}
\end{equation}
In Eq. \eqref{eq:channel1}, 
we are implicitly assuming that the propagation environment is made of $N_{\rm cl}$ scattering clusters, each of which contributes with $N_{{\rm ray}, i}$ propagation paths, $i=1, \ldots, N_{\rm cl}$, plus a  possibly present LOS component.  
We denote by  $\phi_{i,l,k}^{\rm MS}$ and $\phi_{i,l,k}^{\rm BS}$ the downlink angle of arrival  (that coincides with the uplink angle of departure) at the $k$-th MS and the downlink angle of departure (that coincides with the uplink angle of departure) at the BS of the $l^{th}$ ray in the $i^{th}$ scattering cluster for the $k$-th MS, respectively. 
The quantities $\alpha_{i,l,k}$ and $L(r_{i,l,k})$ are the complex path gain and the attenuation associated  to the $(i,l)$-th propagation path of the channel from the BS to the $k$-th MS.
The complex gain  $\alpha_{i,l,k}\thicksim \mathcal{CN}(0, \sigma_{\alpha,i}^2)$, with  $\sigma_{\alpha,i}^2=1$  \cite{spatiallysparse_heath}. The factors $\mathbf{a}_{\rm MS}(\phi_{i,l,k}^{\rm MS})$ and $\mathbf{a}_{\rm BS}(\phi_{i,l,k}^{\rm BS})$ represent the normalized receive and transmit array response vectors evaluated at the corresponding angles of arrival and departure; for a ULA with half-wavelength inter-element spacing we have
$
\mathbf{a}_{\rm BS}(\phi)=\displaystyle \frac{1}{\sqrt{N_{\rm BS}}}[1 \; e^{-j\pi \sin \phi} \; \ldots \; e^{-j\pi (N_{\rm BS}-1) }]^T$.  A similar expression can be also given for $\mathbf{a}_{\rm MS}(\phi_{i,l,k}^{\rm MS})$.
Finally, $\gamma=\displaystyle\sqrt{\frac{N_{\rm BS} N_{\rm MS}}{\sum_{i=1}^{N_{\rm cl}}N_{{\rm ray},i}}}$  is a normalization factor ensuring that the received signal power scales linearly with the product $N_{\rm BS} N_{\rm MS}$, and accounting for the fact that the number of possible transmitter-receiver path in a multiantenna system is proportional to the product of the number of transmit and receive antennas.
Regarding the LOS component, denoting by 
$\phi_{k,{\rm LOS}}^{\rm MS}$,  $\phi_{k, {\rm LOS}}^{\rm BS}$,
the downlink arrival and departure angles corresponding to the LOS link, we assume that
\begin{equation}
\begin{array}{llll}
\mathbf{H}_{k,{\rm LOS}} & =   
I_{k,{\rm LOS}}(d_k) \sqrt{N_{\rm MS} N_{\rm BS} L(d_k)} e^{j \theta_k} \\ & \times \mathbf{a}_{\rm MS}(\phi_{k,{\rm LOS}}^{\rm MS})  \mathbf{a}_{\rm BS}^H(\phi_{k,{\rm LOS}}^{\rm BS}) \; .
\end{array}
\label{eq:Hlos}
\end{equation}
In the above equation, $\theta_k \thicksim \mathcal{U}(0 ,2 \pi)$, $d_k$ is the distance between the BS and the $k$-th MS, while $I_{k,{\rm LOS}}(d_k) $ is a random variate indicating if a LOS link exists between transmitter and receiver. 
A detailed description of all the parameters needed for the generation of sample realizations for the channel model of Eq. \eqref{eq:channel1} is reported in \cite{buzzidandreachannel_model}, and we refer the reader to this reference for further details on the channel model. We also assume that the BS-to-MS channels $\mathbf{H}_1, \, \mathbf{H}_2, \, \ldots \, 
\mathbf{H}_K$ are statistically independent, a reasonable assumption provided that there are no close MSs and they have random orientations.

\begin{figure*}[t!]
	\centering
    \begin{subfigure}[b]{0.5\textwidth}
    \centering
	\includegraphics[scale=0.6]{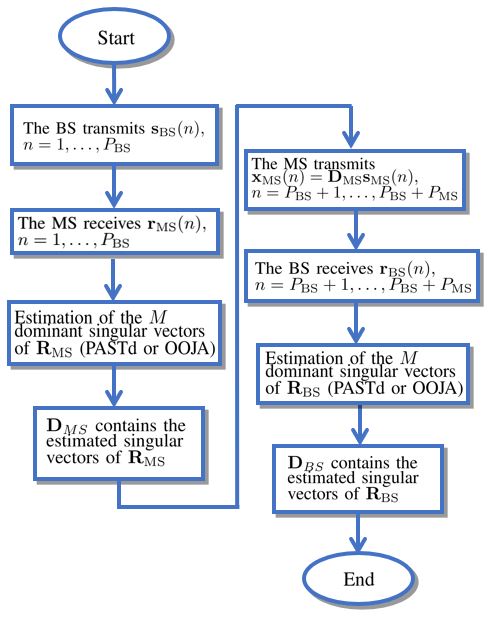}
        \caption{\color{white}.}
    \end{subfigure}%
    \begin{subfigure}[b]{0.5\textwidth}
       \centering
\includegraphics[scale=0.6]{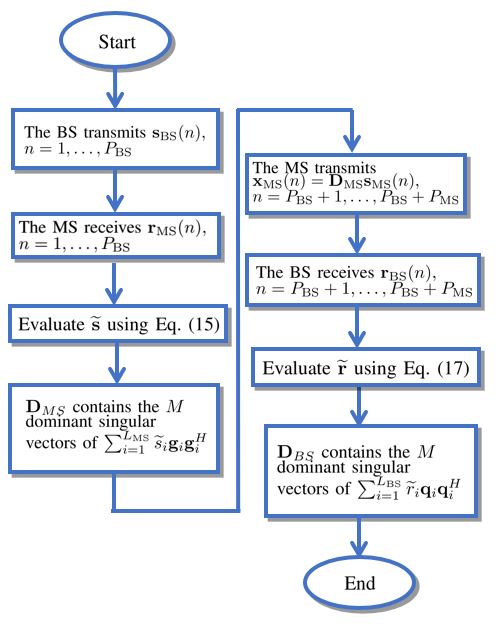}
        \caption{\color{white}.}
    \end{subfigure}
    \caption{Simulation diagrams for the proposed channel estimation algorithms in the single-user scenario. Subspace-based channel estimation in subfigure (a) and LS channel estimation in subfigure (b).}
    \label{Fig:Simulation_Diagram_SU}
\end{figure*}

\section{Subspace-based channel estimation in a single-user scenario} \label{Subspace_single_user_Section}
We start by considering a single BS-to-MS link, that may be  representative either of a single-user wireless link, or of a 
BS - MS link in a wireless cellular system using an orthogonal multiple
access scheme. In the following, we will thus refer to the channel model \eqref{eq:channel1} by omitting, for ease of notation, the index $k$. 
The consideration of the single-user case permits a gentle (and with simplified notation) introduction to the multiuser case, to be considered in the sequel of the paper.
Given the structured (parametric) channel model in \eqref{eq:channel1}, 
and given the use of the TDD protocol, we are actually interested only to the dominant left and right singular vectors of the channel matrix itself; it is also quite easy to realize that these singular vectors tend to coincide, in the limit of large number of antennas, with the ULA array responses at the angles corresponding to the rays associated with  the complex gain with the largest norm.

\subsection{FD beamforming architecture}
We start by considering the case of FD beamforming, i.e. no analog beamforming is performed and the number of RF chains coincides with the number of antennas, both at the BS and at the MS. 
The proposed protocol for channel estimation consists of two successive phases. In  phase (a), the BS transmits a suitable probing signal and the MS estimates the dominant left eigenvectors of the channel matrix; then, in phase (b), the MS transmits a suitable signal and the BS estimates the dominant right eigenvectors of the channel matrix. 

With regard to phase (a), let  $\mathbf{s}_{\rm BS}(n)$, with $n=1, \ldots, P_{\rm BS}$, be a sequence of $N_{\rm BS}$-dimensional random column 
vectors with identity covariance matrix\footnote{As an example, a sequence of random uniform binary-valued antipodal symbols can be used.}. These vectors  are transmitted by the BS at (discrete) time $n=1, \ldots, P_{\rm BS}$; the signal received at the MS at time $n$ is expressed as
the following $(N_{\rm MS} \times 1)$-dimensional vector:

\begin{equation}
\mathbf{r}_{\rm MS}(n)=\mathbf{H}\mathbf{s}_{\rm BS}(n)+\mathbf{w}_{\rm MS}(n) \; ,
\label{eq:receivedMS}
\end{equation} 
where $\mathbf{w}_{\rm MS}(n)$ is the $N_{\rm MS}$-dimensional additive white Gaussian noise (AWGN) vector, modeled as $\mathcal{CN} (0, \sigma^2_n)$ independent random variates (RVs).
Letting  $\mathbf{H}=\mathbf{U}\mathbf{\Lambda}\mathbf{V}^H$ denote the singular value decomposition of the channel matrix,  the covariance matrix of the received signal is expressed as
\begin{equation}
\mathbf{R}_{\rm MS}=\mathbb{E}\left[\mathbf{r}_{\rm MS}(n)\mathbf{r}_{\rm MS}^H(n)\right]=\mathbf{U}\mathbf{\Lambda}^2\mathbf{U}^H+ \sigma^2_n \mathbf{I}_{N_{\rm MS}} \; .
\label{eq:covMS}
\end{equation}
Given \eqref{eq:covMS}, it is thus easily seen that we can estimate the $M$ dominant left singular vectors of the channel matrix by estimating the $M$ dominant directions of the subspace spanned by the received vectors 
$\mathbf{r}_{\rm MS}(n)$, with $n=1, \ldots, P_{\rm BS}$. The signal processing literature is rich of adaptive subspace tracking algorithms that can be straightforwardly applied at the MS to obtain an estimate of the principal left singular vectors of the channel matrix. Deferring later the specification of the adopted subspace tracking algorithms,  let 
$\mathbf{D}_{MS}$ be the $(N_{\rm MS} \times M)$-dimensional matrix containing the estimate of the $M$ dominant singular vectors of $\mathbf{R}_{\rm MS}$; the matrix $\mathbf{D}_{\rm MS}$ will be used at the MS as a precoder during data transmission and as a combiner when receiving data from the BS. 
After $P_{\rm BS}$ symbol intervals, phase (a) is over and phase (b) starts. The MS now transmits random independent vectors with identity covariance matrix in order to enable estimation at the BS of the dominant right eigenvectors of the channel matrix $\mathbf{H}$. More precisely, 
let $\mathbf{s}_{\rm MS}(n)$, with $n=P_{\rm BS}+1, \ldots, P_{\rm BS}+ P_{\rm MS}$, be a sequence of  zero-mean
$M$-dimensional random vectors with identity covariance matrix and assume that the MS transmits  the following $N_{\rm MS}$-dimensional data vectors
\begin{equation}
\mathbf{x}_{\rm MS}(n)=\mathbf{D}_{ \rm MS} \mathbf{s}_{\rm MS}(n) \; .
\end{equation}
The received discrete-time signal at the BS is represented by the following $N_{\rm BS}$-dimensional vector:
\begin{equation}
\mathbf{r}_{\rm BS}(n)=\mathbf{H}^H\mathbf{x}_{\rm MS}(n)+\mathbf{w}_{\rm BS}(n) \; ,
\label{eq:receiveBS}
\end{equation} 
where $\mathbf{w}_{\rm BS}(n)$ is the $N_{\rm BS}$-dimensional AWGN vector, modeled as $\mathcal{CN} (0, \sigma^2_n)$ independent RVs\footnote{Note that in phase (b) the MS is already using the estimated precoder
$\mathbf{D}_{ \rm MS}$; the illustrated procedure also works if the MS does not use the precoder and sends $N_{\rm MS}$-dimensional random vectors.}. 
Similarly to phase (a), under the assumption of negligible errors  in the estimation of the left singular vectors of the channel matrix, the covariance matrix of the received signal at the BS is approximately\footnote{The approximation stems from the fact that we are assuming that $\mathbf{D}_{ \rm MS}\mathbf{D}_{ \rm MS}^H \approx \mathbf{I}_{N_{\rm MS}}.$} expressed as
\begin{equation}
\mathbf{R}_{\rm BS} = 
\mathbb{E}\left[\mathbf{r}_{\rm BS}(n)\mathbf{r}_{\rm BS}^H(n)\right] \approx
\mathbf{V}\mathbf{\Lambda}^2\mathbf{V}^H + \sigma^2_n \mathbf{I}_{N_{\rm BS}} \; ,
\end{equation}
thus implying that the 
 the $M$ dominant right singular vectors of the channel matrix can be estimated by running adaptive subspace tracking algorithms at the BS. We will denote by $\mathbf{D}_{\rm BS}$ is the $(N_{\rm BS} \times M)$-dimensional matrix containing the estimates of the $M$ dominant singular vectors of $\mathbf{R}_{\rm MS}$.

A simulation diagram for the outlined subspace-based channel estimation procedure in the single-user scenario is reported in subfigure (a) of Figure \ref{Fig:Simulation_Diagram_SU}.

\subsection{HY beamforming architecture}
In the previous section a FD beamforming structure has been assumed. We now examine the case in which, for complexity reasons, a HY beamforming architecture is adopted. The front-end processing consists of an analog RF combining matrix aimed at reducing the number of RF chains needed to implement the base-band processing. From a mathematical point of view, 
the beamforming matrices at the MS and at the BS can be expressed as
\begin{equation}
\begin{array}{llll}
\mathbf{D}_{\rm MS}&= & \mathbf{D}_{\rm MS,RF}\mathbf{D}_{\rm MS,BB} \; , \qquad \mbox{and} \\
\mathbf{D}_{\rm BS}&= & \mathbf{D}_{\rm BS,RF}\mathbf{D}_{\rm BS,BB} \; ,
\end{array}
\label{eq:HYcombiners}
\end{equation}
respectively. In \eqref{eq:HYcombiners}, 
 $\mathbf{D}_{\rm MS,RF}$ is an $(N_{\rm MS} \times N_{\rm MS}^{\rm RF})$-dimensional matrix with unit-norm entries, while  $\mathbf{D}_{\rm MS,BB}$ is an $(N_{\rm MS}^{\rm RF} \times M)$-dimensional matrix with no constraint on its entries. Similarly,  $\mathbf{D}_{\rm BS,RF}$ is an $(N_{\rm BS} \times N_{\rm BS}^{\rm RF})$-dimensional matrix with unit-norm entries, and $\mathbf{D}_{\rm BS,BB}$ is an $(N_{\rm BS}^{\rm RF} \times M)$-dimensional baseband combining matrix. 
 The design of HY analog/digital combiners is a vastly explored research topic; most papers try to find the HY combiner that best approximates, according to some criterion, the optimal FD combiner. In this paper, we use a simpler and different approach. We assume that $\mathbf{D}_{\rm MS,RF}$ and $\mathbf{D}_{\rm BS,RF}$ have a fixed structure and in particular contain on their column the ULA array responses corresponding to a grid of discrete angles spanning the range $[-\pi/2, \pi/2]$. In particular, letting 
 \begin{equation}
 \begin{array}{lll}
\vartheta_i^{\rm MS}= \left(-\frac{\pi}{2}+\frac{\pi(i-1)}{N_{\rm MS}^{\rm RF}}\right)\; , \; \; i=1, \ldots, N_{\rm MS}^{\rm RF} \; , \\
\vartheta_i^{\rm BS}= \left(-\frac{\pi}{2}+\frac{\pi(i-1)}{N_{\rm BS}^{\rm RF}}\right)\; , \; \; i=1, \ldots, N_{\rm BS}^{\rm RF} \; ,
 \end{array}
 \end{equation}
 the RF combiners have the following structure:
\begin{equation}
\begin{array}{lll}
\mathbf{D}_{\rm BS,RF}& =& \left[\mathbf{a}_{\rm BS}\left(\vartheta_1^{\rm BS} \right), \ldots \mathbf{a}_{\rm BS}\left(\vartheta_{N_{\rm BS}^{\rm RF}}^{\rm BS}\right) \right] \; , \\
\mathbf{D}_{\rm MS,RF}& =& \left[\mathbf{a}_{\rm MS}\left(\vartheta_1^{\rm MS} \right), \ldots \mathbf{a}_{\rm MS}\left(\vartheta_{N_{\rm MS}^{\rm RF}}^{\rm MS}\right) \right] \; .
\end{array}
\end{equation}
Now, focus on the scheme in Figure \ref{fig:scheme} and consider the  cascade of the BS analog beamformer, the channel $\mathbf{H}$ and the MS analog beamformer; it is straightforward to show that  this cascade can be modeled through the matrix 
$\widetilde{\mathbf{H}}=\mathbf{D}^H_{\rm MS,RF} \mathbf{H} \mathbf{D}_{\rm BS,RF}$, of dimension 
$N_{\rm MS}^{\rm RF} \times N_{\rm BS}^{\rm RF}$. As a consequence, the channel estimation scheme outlined in the previous section under the assumption of FD beamforming can be now re-applied on the (reduced-dimension) composite channel $\widetilde{\mathbf{H}}$.

\subsection{Subspace tracking algorithms}
We now adapt two well-known subspace-tracking algorithms to our context. We start with the PASTd, introduced in \cite{yang1995projection}; one of its most popular applications deals with the design of blind multiuser detection for code-division multiple access system, as detailed in the highly-cited paper \cite{wang1998blind}. In order to illustrate this algorithm, let $\mathbf{r}$ be an $(N \times 1)$-dimensional random vector with autocorrelation matrix $\mathbf{C}=\mathbb{E}\left[\mathbf{r}\mathbf{r}^H\right]$. Consider the scalar function 
\begin{equation}
\begin{array}{lll}
J\left(\mathbf{W}\right)& =\mathbb{E}\left[\|\mathbf{r}-\mathbf{W}\mathbf{W}^T\mathbf{r}\|^2\right] \\ &=\text{tr}(\mathbf{C})-2\text{tr}(\mathbf{W}^T\mathbf{C}\mathbf{W})+\text{tr}(\mathbf{W}^T\mathbf{C}\mathbf{W}\mathbf{W}^T\mathbf{W}),
\end{array}
\label{cost_function}
\end{equation}
with a $(N \times M)$-dimensional matrix argument $\mathbf{W}$, with $M<N$. It is shown in \cite{yang1995projection} that
\begin{itemize}
\item[-] the matrix $\mathbf{W}$ is a stationary point of $J\left(\mathbf{W}\right)$ if and only if 
$\mathbf{W}=\mathbf{T}_M\mathbf{Q}$, where $\mathbf{T}_M$ is a $(N \times M)$-dimensional matrix contains any $M$ distinct eigenvectors of $\mathbf{C}$ and $\mathbf{Q}$ is any $(M \times M)$-dimensional unitary matrix.
\item[-] All stationary points of $J\left(\mathbf{W}\right)$ are saddle points except when $\mathbf{T}_M$ contains the $M$ dominant eigenvectors of $\mathbf{C}$. In that case, $J\left(\mathbf{W}\right)$ attains the global minimum.
\end{itemize}
Therefore, for $M=1$, the solution of minimizing $J\left(\mathbf{W}\right)$ is given by the most dominant eigenvector of $\mathbf{C}$. In practical applications, only sample vectors $\mathbf{r}(i)$ are available and so the statistical average in \eqref{cost_function} is replaced by an exponentially-windowed time-average, i.e.:
\begin{equation}
J\left[\mathbf{W}(n)\right]=\sum_{i=1}^n{\beta^{n-i}\|\mathbf{r}(i)-\mathbf{W}(n)\mathbf{W}(n)^T\mathbf{r}(i)\|^2},
\label{cost_sum}
\end{equation}
where $\beta$ is the forgetting factor. The key trick of the PASTd approach is to approximate $\mathbf{W}(n)^T\mathbf{r}(i)$ in \eqref{cost_sum}, the unknown projection of $\mathbf{r}(i)$ onto the columns of $\mathbf{W}(n)$, by $\mathbf{y}(i)=\mathbf{W}(i-1)^T\mathbf{r}(i)$, which can be calculated for $i=1,\ldots n$ at sampling time $n$. This results in a modified cost function 
\begin{equation}
\tilde{J}\left[\mathbf{W}(n)\right]=\sum_{i=1}^n{\beta^{n-i}\|\mathbf{r}(i)-\mathbf{W}(n)\mathbf{y}(n)\|^2},
\label{modified_cost_sum}
\end{equation}
The recursive least squares (RLS) algorithm can then be used to solve for $\mathbf{W}(n)$ that minimizes the exponentially weighted least squares criterion \eqref{modified_cost_sum}. When there is the need to track the $M$ dominant eigenvectors with $M>1$, the PASTd algorithm adopts the  deflation technique and its basic idea is as follows: for $M=1$, by minimizing $\tilde{J}\left[\mathbf{W}(n)\right]$ in \eqref{modified_cost_sum} the  dominant eigenvector is updated; then, the contribution from this estimated eigenvector is removed from $\mathbf{r}(n)$ itself, and the second most dominant eigenvector can be now extracted from the data. This procedure is applied repeatedly until the $M$ dominant eigenvectors are sequentially estimated.  The complete PASTd procedure to be implemented at the MS for the estimation of the $M$ dominant left singular vectors is reported\footnote{The procedure to be implemented in phase (b) at the BS resembles that carried out in phase (a) and is omitted for brevity.} in Algorithm \ref{PASTd}. 

A simulation diagram for the outlined subspace-based channel estimation procedure in the single-user scenario is reported in subfigure (a) of Figure \ref{Fig:Simulation_Diagram_SU}.

\begin{algorithm}[!t]

\caption{The PASTd Algorithm to be implemented during phase (a) at the MS.}

\begin{algorithmic}[1]

\label{PASTd}
\STATE  Set $\beta=0.995$
\FOR { $n=1:P_{\rm BS}$}

 \STATE $\mathbf{x}_1(n)=\mathbf{r}_{\rm MS}(n)$

\FOR { $m=1:M$}

\STATE  {$y_m(n)=\mathbf{u}_m^H(n-1)\mathbf{x}_m(n)$}

\STATE {$\lambda_m(n)=\beta\lambda_m(n-1)+|y_m(n)|^2$}

\STATE {$\mathbf{u}_m(n)\! \! = \! \! \mathbf{u}_m(\!n-1\!)\!+\!\left[\mathbf{x}_m(n)\!-\!\mathbf{u}_m(n-1)y_m(n)\right]\frac{y_m(n)^*}{\lambda_m(n)}$}

\STATE $\mathbf{x}_{m+1}(n)=\mathbf{x}_m(n)-\mathbf{u}_m(n)y_m(n)$

\ENDFOR

\ENDFOR

\STATE $\mathbf{D}_{\rm MS}=\left[ \mathbf{u}_1(P_{\rm BS}) , \mathbf{u}_2(P_{\rm BS}), \ldots,
\mathbf{u}_M(P_{\rm BS})\right] $  
\end{algorithmic}

\end{algorithm}

\medskip

Another possible subspace tracking algorithm is the OOJA algorithm, that was introduced in \cite{abed2000orthogonal}, building upon the minor subspace extraction algorithm of Oja originally proposed in \cite{oja1992principal}. 
Focusing again on phase (a), and letting $\mathbf{W}(n)$ denote the estimate of the beamformer $\mathbf{D}_{MS}$ available at time epoch $n$, the complete algorithm' recipe to be run at the MS is reported in  Algorithm \ref{OOJA_Algorithm}.

\begin{algorithm}[!t]

\caption{The OOJA Algorithm}

\begin{algorithmic}[1]

\label{OOJA_Algorithm}
\STATE  Set $\delta=0.01$

\FOR { $n=1:P_{\rm BS}$}

\STATE $\mathbf{v}(n)=\mathbf{W}^T(n)\mathbf{r}_{\rm MS}(n)$

\STATE  $\mathbf{z}(n)=\mathbf{W}(n)\mathbf{v}(n)$

\STATE  $\mathbf{p}(n)=\mathbf{r}_{\rm MS}(n)-\mathbf{z}(n)$

\STATE  $\varphi(n)=\frac{1}{\sqrt{1+\delta^2\|\mathbf{p}(n)\|^2\|\mathbf{v}(n)\|^2}}$

\STATE  $\tau(n)=\frac{\phi(n)-1}{\|\mathbf{v}(n)\|^2}$

\STATE  $\bar{\mathbf{p}}(n)=-\frac{\tau(n)\mathbf{z}(n)}{\delta}+\phi(n)\mathbf{p}(n)$

\STATE  $\mathbf{W}(n+1)=\mathbf{W}(n)-\delta\bar{\mathbf{p}}(n)\mathbf{v}^T(n)$

\ENDFOR

\STATE $\mathbf{D}_{\rm MS}=\mathbf{W}(P_{\rm BS}+1)$  

\end{algorithmic}

\end{algorithm}

\section{LS channel estimation in the single-user scenario} \label{LS_single_user_section}

We now exploit an LS approach  \cite{kay1998}.

\subsection{FD beamforming architecture}

We start by considering the case of FD beamforming. Also in the case, the proposed protocol consists of two successive phases. In  phase (a), the BS transmits a suitable probing signal and the MS estimates the dominant left eigenvectors of the channel matrix; then, in phase (b), the MS transmits a suitable signal and the BS estimates the dominant right eigenvectors of the channel matrix. 

With regard to phase (a), the received signal at the MS is again expressed as \eqref{eq:receivedMS}. Let $\mathbf{E}_{\rm MS}=\frac{1}{P_{\rm BS}}\sum_{n=1}^{P_{\rm BS}} {\mathbf{r}_{\rm MS}(n)\mathbf{r}_{\rm MS}(n)^H}$ be the sample covariance matrix of the noisy received vectors $\mathbf{r}_{\rm MS}(n) \; , n=1, \ldots , P_{\rm BS}$. 
Due to the clustered nature of the mmWave channel matrix, the covariance matrix of the data at the MS can be expressed as the sum of rank-1 matrices whose principal eigevector is the MS array response corresponding to the directions 
$\left\{\phi_{i,l}^{\rm MS}\right\}_{l=1, \ldots, N_{{\rm ray},i} \, i=1, \ldots, N_{\rm cl}}$. 
Accordingly, letting $\theta_i=\frac{2 \pi (i-1)}{L_{\rm MS}} \; , i=1, \ldots , L_{\rm MS} $ denote a set of  $L_{\rm MS} \gg N_{\rm MS}$ angles uniformly spanning the entire $2\pi$ angular range, and denoting by $\mathbf{g}_i$ the MS array response corresponding at the angle $\theta_i$, we aim at approximating the matrix $\mathbf{E}_{\rm MS}$ with 
 the sum $\sum_{i=1}^{L_{\rm MS}} {s_i \mathbf{g}_i \mathbf{g}_i^H}$, where the coefficients $s_1, \ldots, s_{L_{\rm MS}}$ are unknown positive real numbers, that can be determined by considering 
 the following LS problem
\begin{equation}
\widetilde{\mathbf{s}}= \underset{s_1, \ldots, s_{L_{\rm MS}}}{\arg\min} \; \left\| \sum_{i=1}^{L_{\rm MS}} {s_i \mathbf{g}_i \mathbf{g}_i^H} - \mathbf{E}_{\rm MS} \right\|^2 \; .
\label{LS_problem}
\end{equation}
Solving \eqref{LS_problem}  leads to the solution
\begin{equation}
\widetilde{\mathbf{s}}=\mathbf{F}_{\rm MS}^{-1} \mathbf{e}_{\rm MS},
\label{eq:solutionLS}
\end{equation}
where $\left[\mathbf{F}_{\rm MS}\right]_{i,j}=\text{tr} \left( \mathbf{g}_i \mathbf{g}_i^H \mathbf{g}_j \mathbf{g}_j^H\right)$ and $\left[\mathbf{e}_{\rm {MS}}\right]_i=\mathfrak{R}\left\lbrace\text{tr} \left( \mathbf{E}_{\rm MS} \mathbf{g}_i \mathbf{g}_i^H\right) \right\rbrace$.
After computing the vector $\widetilde{\mathbf{s}}$ by using \eqref{eq:solutionLS}, the MS beamformer is obtained by 
retaining the $M$ dominant eigenvectors 
of the matrix $\sum_{i=1}^{L_{\rm MS}} {\widetilde{s}_i \mathbf{g}_i \mathbf{g}_i^H}$.

After $P_{\rm BS}$ symbol intervals, phase (a) is over and phase (b) starts. 
 Similarly to the case of subspace-based channel estimation, the received discrete-time signal at the BS is represented by Eq. \eqref{eq:receiveBS}. Let $\mathbf{E}_{\rm BS}=\frac{1}{P_{\rm MS}}\sum_{n=1}^{P_{\rm MS}} {\mathbf{r}_{\rm BS}(n)\mathbf{r}_{\rm BS}(n)^H}$ be the sample covariance matrix of the noisy received vectors $\mathbf{r}_{\rm BS}(n) \; , n=P_{\rm BS}+1, \ldots , P_{\rm BS}+P_{\rm MS}$. Again, we can write the matrix $\mathbf{E}_{\rm BS}$ through the  approximation
$\mathbf{E}_{\rm BS} \approx \sum_{i=1}^{L_{\rm BS}} {r_i \mathbf{q}_i \mathbf{q}_i^H}$, 
where $\mathbf{q}_i=\mathbf{a}_{\rm BS} \left(\phi^{\rm BS}_i \right)$, and $\phi^{\rm BS}_i=\frac{2 \pi (i-1)}{L_{\rm BS}} \; , i=1, \ldots , L_{\rm BS} $, with $ L_{\rm BS} \gg N_{\rm BS}$.
In order to estimate the right eigenvectors of the channel matrix we formulate the following LS problem:
\begin{equation}
\widetilde{\mathbf{r}}= \underset{\mathbf{r}}{\arg\min} \; \| \sum_{i=1}^{L_{\rm BS}} {r_i \mathbf{q}_i \mathbf{q}_i^H} - \mathbf{E}_{\rm BS} \|^2 \; ,
\label{LS_problem_2}
\end{equation}
whose solution is shown to be expressed as 
\begin{equation}
\widetilde{\mathbf{r}}=\mathbf{F}_{\rm BS}^{-1} \mathbf{e}_{\rm BS},
\label{eq:solutionLS2}
\end{equation}
where now $\left[\mathbf{F}_{\rm BS}\right]_{i,j}=\text{tr} \left( \mathbf{q}_i \mathbf{q}_i^H \mathbf{q}_j \mathbf{q}_j^H\right)$ and  $\left[\mathbf{e}_{\rm {BS}}\right]_i=\mathfrak{R}\left\lbrace\text{tr} \left( \mathbf{E}_{\rm BS} \mathbf{q}_i \mathbf{q}_i^H\right) \right\rbrace$.
After computing the vector $\widetilde{\mathbf{r}}$ by using \eqref{eq:solutionLS2}, the BS beamformer 
the matrix $\mathbf{D}_{\rm BS}$ is built by 
retaining the $M$ dominant eigenvectors 
of the matrix $\sum_{i=1}^{L_{\rm BS}} {\widetilde{r}_i \mathbf{q}_i \mathbf{q}_i^H}$.

A simulation diagram for the outlined LS-based channel estimation procedure in the single-user scenario is reported in subfigure (b) of Figure \ref{Fig:Simulation_Diagram_SU}.

\begin{figure*}[t!]
	\centering
    \begin{subfigure}[b]{0.5\textwidth}
    \centering
	\includegraphics[scale=0.6]{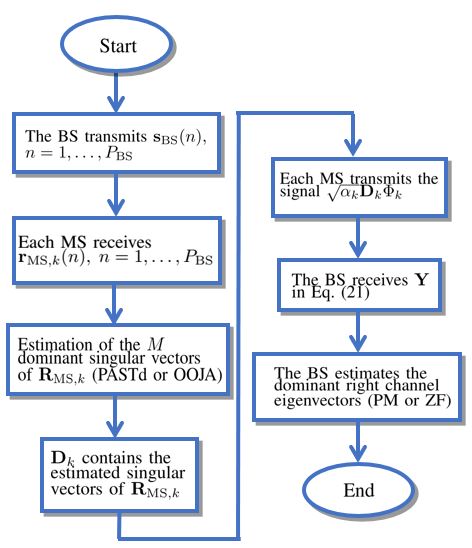}
        \caption{\color{white}.}
    \end{subfigure}%
    \begin{subfigure}[b]{0.5\textwidth}
       \centering
\includegraphics[scale=0.6]{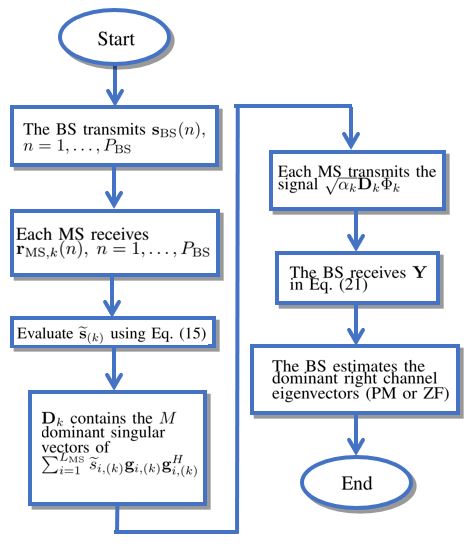}
        \caption{\color{white}.}
    \end{subfigure}
    \caption{Simulation diagrams for the proposed channel estimation algorithms in the multiuser scenario. Subspace-based channel estimation in subfigure (a) and LS channel estimation in subfigure (b).}
    \label{Fig:Simulation_Diagram_MU}
\end{figure*}

\subsection{HY beamforming architecture}

We now extend the LS-based channel estimation scheme to the case in which a HY beamforming architecture is adopted. 
We thus assume that the beamforming matrices at the MS and at the BS can be expressed as in \eqref{eq:HYcombiners},
where  $\mathbf{D}_{\rm MS,RF}$ and $\mathbf{D}_{\rm BS,RF}$ have a fixed structure and in particular contain on their column the ULA array responses corresponding to a grid of discrete angles spanning the range $[-\pi/2, \pi/2]$
with quantization step given by $\pi$ divided by the number of considered RF chains. 
We can consider the composite channel 
$\widetilde{\mathbf{H}}=\mathbf{D}^H_{\rm MS,RF} \mathbf{H} \mathbf{D}_{\rm BS,RF}$ and apply to it the LS procedure.  
With regard to phase (a), let  $\tilde{\mathbf{s}}_{\rm BS}(n)$, with $n=1, \ldots, P_{\rm BS}$, be a sequence of $N_{\rm BS}^{\rm RF}$-dimensional random column 
vectors with identity covariance matrix. These vectors  are transmitted by the BS at (discrete) time $n=1, \ldots, P_{\rm BS}$; the signal received at the MS at time $n$ is expressed as
the following $(N_{\rm MS}^{\rm RF} \times 1)$-dimensional vector:
\begin{equation}
\widetilde{\mathbf{r}}_{\rm MS}(n)=\widetilde{\mathbf{H}}\widetilde{\mathbf{s}}_{\rm BS}(n)+\mathbf{D}_{\rm MS,RF}^H\widetilde{\mathbf{w}}_{\rm MS}(n) \; ,
\label{eq:receivedMS_LS_HY}
\end{equation}
 where $\widetilde{\mathbf{w}}_{\rm MS}(n)$ is the $N_{\rm MS}^{\rm RF}$-dimensional AWGN vector, modeled as $\mathcal{CN} (0, \sigma^2_n)$ independent RVs. Let $\widetilde{\mathbf{E}}_{\rm MS}=\frac{1}{P_{\rm BS}}\sum_{n=1}^{P_{\rm BS}} {\widetilde{\mathbf{r}}_{\rm MS}(n)\widetilde{\mathbf{r}}_{\rm MS}(n)^H}$ be the sample covariance matrix of the noisy received vectors $\widetilde{\mathbf{r}}_{\rm MS}(n) \; , n=1, \ldots , P_{\rm BS}$. Similarly to the case of LS channel estimation with FD beamforming we can approximate the sample covariance matrix as
 $
\widetilde{\mathbf{E}}_{\rm MS} \approx \sum_{i=1}^{L_{\rm MS}} {{s}_i \widetilde{\mathbf{g}}_i \widetilde{\mathbf{g}}_i^H}$, where now 
 $\widetilde{\mathbf{g}}_i=\mathbf{D}^H_{\rm MS,RF}\mathbf{a}_{\rm MS} \left(\phi^{\rm MS}_i \right)$, and $\phi^{\rm MS}_i=\frac{2 \pi (i-1)}{L_{\rm MS}} \; , i=1, \ldots , L_{\rm MS} $, with $ L_{\rm MS} \gg N_{\rm MS}$.
We formulate the LS problem
\begin{equation}
\widetilde{\mathbf{s}}= \underset{{\mathbf{s}}}{\arg\min} \; \| \sum_{i=1}^{L_{\rm MS}} {{s}_i \widetilde{\mathbf{g}}_i \widetilde{\mathbf{g}}_i^H} - \widetilde{\mathbf{E}}_{\rm MS} \|^2,
\label{LS_problem_hybrid}
\end{equation}
with solution $\widetilde{\mathbf{s}}=\widetilde{\mathbf{F}}_{\rm MS}^{-1} \widetilde{\mathbf{e}}_{\rm MS}$, where 
 $\left[\widetilde{\mathbf{F}}_{\rm MS}\right]_{i,j}=\text{tr} \left( \widetilde{\mathbf{g}}_i \widetilde{\mathbf{g}}_i^H \widetilde{\mathbf{g}}_j \widetilde{\mathbf{g}}_j^H\right)$ and  $\left[\widetilde{\mathbf{e}}_{\rm {MS}}\right]_i=\mathfrak{R}\left\lbrace\text{tr} \left( \widetilde{\mathbf{E}}_{\rm MS} \widetilde{\mathbf{g}}_i \widetilde{\mathbf{g}}_i^H\right) \right\rbrace$. The precoder 
$\mathbf{D}_{\rm MS, BB}$ is then obtained by considering the  $M$ dominant singular vectors of 
the matrix $\sum_{i=1}^{L_{\rm MS}} {\widetilde{s}_i \widetilde{\mathbf{g}}_i \widetilde{\mathbf{g}}_i^H} $.

For phase (b), a similar procedure is pursued. 
Let  $\tilde{\mathbf{s}}_{\rm MS}(n)$, with $n=P_{\rm BS}+1, \ldots, P_{\rm BS}+P_{\rm MS}$, be a sequence of $M$-dimensional random column 
vectors with identity covariance matrix. These vectors  are transmitted by the MS at (discrete) time $n=P_{\rm BS}+1, \ldots, P_{\rm BS}+P_{\rm MS}$ ; the signal received at the BS at time $n$ is expressed as
the following $(N_{\rm BS}^{\rm RF} \times 1)$-dimensional vector:
\begin{equation}
\widetilde{\mathbf{r}}_{\rm BS}(n)=\widetilde{\mathbf{H}}^H \mathbf{D}_{\rm MS, BB}\widetilde{\mathbf{s}}_{\rm MS}(n)+\mathbf{D}_{\rm BS,RF}^H\widetilde{\mathbf{w}}_{\rm BS}(n) \; ,
\label{eq:receivedBS_LS_HY}
\end{equation}
 where $\widetilde{\mathbf{w}}_{\rm MS}(n)$ is the $N_{\rm BS}^{\rm RF}$-dimensional AWGN vector, modeled as $\mathcal{CN} (0, \sigma^2_n)$ independent RVs. Let $\widetilde{\mathbf{E}}_{\rm BS}=\frac{1}{P_{\rm MS}}\sum_{n=1}^{P_{\rm MS}} {\widetilde{\mathbf{r}}_{\rm BS}(n)\widetilde{\mathbf{r}}_{\rm BS}(n)^H}$ be the sample covariance matrix of the noisy received vectors $\widetilde{\mathbf{r}}_{\rm BS}(n) \; , n=P_{\rm BS}+1, \ldots , P_{\rm BS}+P_{\rm MS}$. 
The  $(N_{\rm BS}^{\rm RF} \times M)$-dimensional matrix  $\mathbf{D}_{\rm BS, BB}$ is built with the $M$ dominant singular vectors of the matrix $\sum_{i=1}^{L_{\rm BS}} {\widetilde{r}_i \widetilde{\mathbf{q}}_i \widetilde{\mathbf{q}}_i^H}$, with 
  $\widetilde{\mathbf{q}}_i=\mathbf{D}^H_{\rm BS,RF}\mathbf{a}_{\rm BS} \left(\phi^{\rm BS}_i \right)$,  $\phi^{\rm BS}_i=\frac{2 \pi (i-1)}{L_{\rm BS}} \; , i=1, \ldots , L_{\rm BS} $,  $ L_{\rm BS} \gg N_{\rm BS}$, and the coefficients 
  $\widetilde{r}_i$ are the entries of the vector 
$\widetilde{\mathbf{r}}=\widetilde{\mathbf{F}}_{\rm BS}^{-1} \widetilde{\mathbf{e}}_{\rm BS}$,  with 
$\left[\widetilde{\mathbf{F}}_{\rm BS}\right]_{i,j}=\text{tr} \left( \widetilde{\mathbf{q}}_i \widetilde{\mathbf{q}}_i^H \widetilde{\mathbf{q}}_j \widetilde{\mathbf{q}}_j^H\right)$ and  $\left[\widetilde{\mathbf{e}}_{\rm {BS}}\right]_i=\mathfrak{R}\left\lbrace\text{tr} \left( \widetilde{\mathbf{E}}_{\rm BS} \widetilde{\mathbf{q}}_i \widetilde{\mathbf{q}}_i^H\right) \right\rbrace$.

\section{Multiuser channel estimation}
While previous derivations considered a single-user link, we now tackle the multiuser scenario;
 we thus focus on a single-cell wireless network wherein a BS serves $K$ MSs in the same time-frequency resource. As in the single-user scenario, we develop a framework that consists of two successive phases. In  phase (a), the BS sends a suitable probing signal in order to let the $K$ MSs estimate the dominant left eigenvectors of their own channel matrix; then, in phase (b), the MSs, using the estimated vectors as pre-coding beamformers, simultaneously send pilot sequences to enable channel estimation at the BS. We first outline the subpace-based technique and then focus on LS channel estimation.
Simulation diagrams for the outlined subspace-based and LS-based channel estimation procedures in the multiuser case are reported in subfigures (a) and (b) of Figure \ref{Fig:Simulation_Diagram_MU}, respectively.

\subsection{Subspace-based channel eigendirections estimation} \label{Multiuser_subspace_section}

\subsubsection{FD architecture}

As in the single-user scenario, we start by considering the case of FD beamforming, i.e. no analog beamforming is performed and the number of RF chains coincides with the number of antennas, both at the BS and at the MSs.  

With regard to phase (a), the BS transmits a suitable probing signal and the MSs estimate the dominant left eigenvectors of their respective BS-to-MS channel matrix. The processing needed in this phase at each MS is exactly coincident with that outlined in the single-user case in Section \ref{Subspace_single_user_Section}, since only the BS transmits and each MS receives an interference-free signal. 
Let thus
$\mathbf{D}_{k}$ be the $(N_{\rm MS} \times M)$-dimensional matrix containing an estimate of the $M$ dominant singular vectors of the covariance matrix of the $N_{MS}$-dimensional vector received
at the $k$-th MS.

After $P_{\rm BS}$ symbol intervals, phase (a) is over and phase (b) starts. Let us now denote by $\mathbf{\Phi}_k$ an $(M \times P_{\rm MS})$-dimensional matrix containing on its rows the $M$  unit-energy pilot sequences assigned to user $k$; we assume that the rows of  $\mathbf{\Phi}_k$ are orthogonal, while no orthogonality is required for pilot sequences assigned to different users. In particular, we will use in the numerical simulations random binary pilot sequences with the constraint that the rows of each matrix $\mathbf{\Phi}_k$ be orthogonal.
The generic $k$-th MS transmits, over $P_{\rm MS}$ consecutive signaling slots, the  matrix 
$\sqrt{\alpha_k} \mathbf{D}_{k} \Phi_k$, with
$\alpha_k$  a proper coefficient ruling the power transmitted by the $k$-th MS. The signal received at the BS can be thus expressed as the following 
$(N_{\rm BS}\times P_{\rm MS})$-dimensional matrix:
\begin{equation}
\begin{array}{lll}
\mathbf{Y}&\! = \! \displaystyle \sum_{k=1}^K \sqrt{\alpha_k}{\mathbf{H}}_k^H \mathbf{D}_{k} \Phi_k \!+ \mathbf{Z} 
& \!\approx \!\displaystyle \sum_{k=1}^K \sum_{i=1}^M\! \lambda_{k,i}\mathbf{v}_{k,i}\Phi_k(i,:) \!+ \mathbf{Z}
\end{array}
\label{eq:received_BS}
\end{equation}
wherein  $\mathbf{Z}$ contains the thermal noise contribution, we have assumed, with no loss of generality, that the diagonal entries of $\mathbf{\Lambda}_k$ are ordered according to a decreasing magnitude, and $\lambda_{k,i}$ and $\mathbf{v}_{k,i}$ represent the $i$-th diagonal entry of $\sqrt{\alpha_k} \mathbf{\Lambda}_k$ and the $i$-th column of $\mathbf{V}_k$, respectively.
Now, based on observable \eqref{eq:received_BS}, and relying on the knowledge of the pilot matrices $\Phi_k$, a number of algorithms can be envisaged to estimate the dominant right channel eigenvectors $\mathbf{v}_{k,i}$. 
A simple algorithm relies on pilot-matched (PM) filtering, i.e. the product $\lambda_{k,i} \mathbf{v}_{k,i}$ can be estimated as follows:
$\widehat{\lambda_{k,i}  \mathbf{v}}_{k,i}=\mathbf{Y}[\mathbf{\Phi}_k(i,:)]^H$.
Alternatively, if  $P_{\rm MS}\geq MK$,  a zero-forcing (ZF) approach can be also used; in particular, the estimate of the matrix
$[\lambda_{k,1} \mathbf{v}_{k,1}, \ldots,  \lambda_{k,M} \mathbf{v}_{k,M}]$ if formed by considering the statistic $\mathbf{Y}\mathbf{Z}_k$, where the $(P_{\rm MS} \times M)$-dimensional matrix $\mathbf{Z}_k$ is such that $\mathbf{\Phi}_k \mathbf{Z}_k=\mathbf{I}_M$ and $\mathbf{\Phi}_j  \mathbf{Z}_k$ is the all-zero matrix for all $j \neq k$. 

\subsubsection{HY architecture}
In the HY architecture beamforming at the BS and at the MSs is of the HY type, so 
the front-end processing both at the BS and at the MSs consists of an analog RF combining matrix aimed at reducing the number of RF chains needed to implement the base-band processing. As in the single-user scenario, the beamforming matrices at the $k$-th MS and at the BS can be expressed as
\begin{equation}
\begin{array}{llll}
\mathbf{D}_{ k}&= & \mathbf{D}_{k,{\rm RF}}\mathbf{D}_{k, {\rm BB}} \; , \qquad \mbox{and} \\
\mathbf{D}_{\rm BS}&= & \mathbf{D}_{\rm BS,RF}\mathbf{D}_{\rm BS,BB} \; ,
\end{array}
\label{eq:HYcombiners_multiuser}
\end{equation}
respectively. In \eqref{eq:HYcombiners}, 
 $\mathbf{D}_{k,{\rm RF}}$ is an $(N_{\rm MS} \times N_{\rm MS}^{\rm RF})$-dimensional matrix with unit-norm entries, while  $\mathbf{D}_{k,{\rm BB}}$ is an $(N_{\rm MS}^{\rm RF} \times M)$-dimensional matrix with no constraint on its entries. Similarly,  $\mathbf{D}_{\rm BS,RF}$ is an $(N_{\rm BS} \times N_{\rm BS}^{\rm RF})$-dimensional matrix with unit-norm entries, and $\mathbf{D}_{\rm BS,BB}$ is an $(N_{\rm BS}^{\rm RF} \times M)$-dimensional baseband combining matrix. As in the single-user case, we assume that $\mathbf{D}_{k,{\rm RF}}$ and $\mathbf{D}_{\rm BS,RF}$ have a fixed structure and in particular contain on their columns the ULA array responses corresponding to a grid of discrete angles uniformly spanning the range $[-\pi/2, \pi/2]$. 
In particular, letting 
 \begin{equation}
 \begin{array}{lll}
\varphi_i^{{\rm MS}, k}= \left(-\frac{\pi}{2}+\frac{\pi(i-1)}{N_{\rm MS}^{\rm RF}}\right)\; , \; \; i=1, \ldots, N_{\rm MS}^{\rm RF} \; , \\
\varphi_i^{\rm BS}= \left(-\frac{\pi}{2}+\frac{\pi(i-1)}{N_{\rm BS}^{\rm RF}}\right)\; , \; \; i=1, \ldots, N_{\rm BS}^{\rm RF} \; ,
 \end{array}
 \end{equation}
 the RF combiners have the following structure:
\begin{equation}
\begin{array}{lll}
\mathbf{D}_{\rm BS,RF}& =& \left[\mathbf{a}_{\rm BS}\left(\varphi_1^{\rm BS} \right), \ldots \mathbf{a}_{\rm BS}\left(\varphi_{N_{\rm BS}^{\rm RF}}^{\rm BS}\right) \right] \; , \\
\mathbf{D}_{k,{\rm RF}}& =& \left[\mathbf{a}_{\rm MS}\left(\varphi_1^{{\rm MS}, k} \right), \ldots \mathbf{a}_{\rm MS}\left(\varphi_{N_{\rm MS}^{\rm RF}}^{{\rm MS}, k}\right) \right] \; .
\end{array}
\end{equation}
Now, the cascade of the BS analog beamformer, the channel $\mathbf{H}_k$ and the $k$-th MS analog beamformer can be modeled through the matrix  $\widetilde{\mathbf{H}}_k=\mathbf{D}^H_{k,{\rm RF}} \mathbf{H}_k \mathbf{D}_{\rm BS,RF}$, of dimension 
$N_{\rm MS}^{\rm RF} \times N_{\rm BS}^{\rm RF}$. As a consequence, the channel estimation scheme outlined above under the assumption of FD beamforming can be now re-applied on the (reduced-dimension) composite channels  $\widetilde{\mathbf{H}}_k$, with $k=1, \ldots, K$. 
In particular, the generic $k$-th MS transmits, over $P_{\rm MS}$ consecutive signaling slots, the  matrix $\sqrt{\alpha_k} \mathbf{D}_{k,{\rm BB}} \Phi_k$, with
$\alpha_k$, again a proper coefficient ruling the power transmitted by the $k$-th MS. The signal received at the BS can be thus expressed as the following 
$(N_{\rm BS}^{\rm RF} \times P_{\rm MS})$-dimensional matrix:
\begin{equation}
\begin{array}{lll}
\mathbf{\widetilde{Y}}&\! = \! \displaystyle \sum_{k=1}^K \sqrt{\alpha_k}\widetilde{\mathbf{H}}_k^H \mathbf{D}_{k,{\rm BB}} \Phi_k \!+ \mathbf{Z} 
& \!\approx \!\displaystyle \sum_{k=1}^K \sum_{i=1}^M\! \widetilde{\lambda}_{k,i}\mathbf{\widetilde{v}}_{k,i}\Phi_k(i,:) \!+ \mathbf{Z}
\end{array}
\label{eq:received_BS_hybrid}
\end{equation}
wherein  $\mathbf{Z}$ contains the thermal noise contribution. Now, based on observable \eqref{eq:received_BS_hybrid}, again either PM or ZF channel estimation can be applied at the BS to estimate the channel eigendirections from the active users. 

Notice also that letting $N_{\rm BS}^{\rm RF}=N_{\rm BS}$ and $N_{\rm MS}^{\rm RF}=N_{\rm MS}$  and taking the analog beamformers equal to an identity matrix the procedures developed in the following describe a system with FD  beamforming.

\subsection{LS channel estimation} \label{Multiuser_LS_section}

\subsubsection{FD architecture}

Again, we start by considering the case of FD beamforming, i.e. no analog beamforming is performed and the number of RF chains coincides with the number of antennas, both at the BS and at the MSs.  

With regard to phase (a), the BS transmits a suitable probing signal and the MSs estimate the dominant left eigenvectors of their respective BS-to-MS channel matrix. 
Also in this case, the processing needed at each MS coincides with that outlined in the single-user case in Section \ref{LS_single_user_section}. 
Let thus
$\mathbf{D}_{k,{\rm BB}}$ be the $(N_{\rm MS}^{\rm RF} \times M)$-dimensional matrix containing the estimate, available at the $k$-th MS,  of the $M$ dominant singular vectors of $\mathbf{H}_{k}$. 

After $P_{\rm BS}$ symbol intervals, phase (a) is over and phase (b) starts. 
Following the same approach as in Section \ref{Multiuser_subspace_section}, the signal received at the BS can be thus expressed as in Eq. \eqref{eq:received_BS}, and either PM or ZF channel estimation can be pursued.

\subsubsection{HY architecture}
As in Section \ref{Multiuser_subspace_section}, the beamforming matrices at the $k$-th MS and at the BS can be expressed as Eq. \eqref{eq:HYcombiners_multiuser} and the cascade of the BS analog beamformer, the channel $\mathbf{H}_k$ and the $k$-th MS analog beamformer can be modeled through the matrix  $\widetilde{\mathbf{H}}_k$. The LS channel estimation scheme outlined in Section \ref{LS_single_user_section} under the assumption of single user can be now re-applied to estimate the left eigendirections of the reduced dimensions channels $\widetilde{\mathbf{H}}_k$ for $k=1, \ldots, K$. After this procedure we obtain $\mathbf{D}_{k,{\rm BB}}$ be the matrix containing the estimate of the $M$ dominant left singular vectors of $\widetilde{\mathbf{H}}_{k}$. 

After $P_{\rm BS}$ symbol intervals, phase (a) is over and phase (b) starts. Similarly to Section \ref{Multiuser_subspace_section}, based on observable \eqref{eq:received_BS_hybrid}, either PM or ZF channel estimation can be applied at the BS to estimate the channel eigendirections from the active users. 

\section{Performance analysis and algorithms comparison}

\subsection{Algorithms comparison}
Before illustrating the performance analysis, it is useful to give some comments about the computational complexity of the proposed subspace-based and LS-based channel estimation algorithms. 

Denoting by $M$ the multiplexing order (i.e., the number of dominant eigenvectors to be tracked) and by $N$ the size of the data-vectors to be processed by the algorithm, 
executing the PASTd algorithm costs $4NM + O(N)$ flops per iteration. Focusing on the single-user case, this implies that the complexity at the BS  is proportional to the product $4N_{\rm BS} M$, for FD beamforming, and to the product $4N_{\rm BS}^{\rm RF} M$, for HY beamforming, respectively. Similarly, at the MS, the complexity  is proportional to the product $4N_{\rm MS} M$, for FD beamforming, and to the product $4N_{\rm MS}^{\rm RF} M$, for HY beamforming, respectively.
Let us now consider the OOJA algorithm, which costs  $3NM + O(N)$ flops per iteration.
Comparing this to the PASTd algorithm, we see that the complexity is still linear in the product between the dimensionality of the processed vectors and the number of principal directions to be tracked, but, compared to the PASTd algorithm, there is a  25$\%$ saving. 
Let us now examine the LS-based channel estimation technique. Differently from the subspace-based techniques, the LS-based algorithm performs a batch (rather than iterative) processing of the data, and the dominant task, from the point of view of the computational complexity, are the matrix inversions in \eqref{eq:solutionLS} and in \eqref{eq:solutionLS2}, and the subsequent SVD computations. Overall, we can thus state that for the LS-based technique the complexity at the BS is $L_{\rm BS}^3 + N_{\rm BS}^3$, for the case of FD beamforming, and $L_{\rm BS}^3 + (N_{\rm BS}^{\rm RF})^3$, for the case of HY beamforming. The complexity at the MS, instead, is  $L_{\rm MS}^3 + N_{\rm MS}^3$, for the case of FD beamforming, and $L_{\rm MS}^3 + (N_{\rm MS}^{\rm RF})^3$, for the case of HY beamforming.
The cubic complexity incurred by the LS-based technique could lead to the conclusion that this algorithm is the most complex one. However, as already discussed, the LS-based algorithm complexity refers to the batch algorithm; thus, in order to perform a fair comparison with the subspace-based methods, the batch complexity must be divides by the length of the pilots, i.e. $P_{\rm BS}$ and $P_{\rm MS}$, for the BS and for the MS algorithms, respectively. This implies that the LS-based algorithm has approximately the same complexity at the subspace-based algorithms, and that  the exact ranking among the algorithms in terms of complexity depends on the particular choice of the involved parameters. 
On the other hand, although this issue is not explored in this paper for the sake of brevity, the subspace-based algorithms should be superior to the LS-based algorithm in terms of tracking capabilities, i.e. they should be able to cope with the situation in which the channel principal directions are subject to (slow) variations.

\subsection{The considered performance measures}
We now describe the performance measures that we will consider to assess the merits of the proposed algorithms.
First of all, we consider the normalized correlation between the true channel eigenvectors and the estimated ones. In particular, we denote by $\mathbf{u}$ and  $\mathbf{v}$ the left and the right eigenvector of the channel matrix corresponding to the dominant eigenvalue respectively, and by $\mathbf{\hat{u}}$ and  $\mathbf{\hat{v}}$ the estimates of left and right dominant eigenvector of the channel matrix, respectively. The normalized correlations are defined as
\begin{equation}
\eta_U=\frac{\left|\mathbf{u}^H\mathbf{\hat{u}}\right|}{\|\mathbf{u}\| \|\mathbf{\hat{u}}\|},
\end{equation}
and as
\begin{equation}
\eta_V=\frac{\left|\mathbf{v}^H\mathbf{\hat{v}}\right|}{\|\mathbf{v}\| \|\mathbf{\hat{v}}\|}.
\end{equation}

The second performance measure here considered is the achievable spectral efficiency in the single user and in the multiuser scenarios. In particular, the achievable spectral efficiency in the single user scenario in downlink and uplink are respectively
\begin{equation}
\begin{array}{lll}
&\mathcal{R}_{\rm MS}=  \log_2 \det \left[ \mathbf{I}_M  \right. \\ & \left. + \ds \frac{P_{\rm T, BS}}{M}\left(\sigma^2_n \mathbf{D}_{\rm MS}^H\mathbf{D}_{\rm MS} \right)^{-1}\mathbf{D}_{\rm MS}^H \mathbf{H}\mathbf{D}_{\rm BS}\mathbf{D}_{\rm BS}^H
\mathbf{H}^H\mathbf{D}_{\rm MS}\right] \; ,
\end{array}
\label{eq:ASE_DL}
\end{equation}
and
\begin{equation}
\begin{array}{lll}
&\mathcal{R}_{\rm BS}=  \log_2 \det \left[ \mathbf{I}_M \right. \\ &+ \left.\ds \frac{P_{\rm T, MS}}{M}\left(\sigma^2_n \mathbf{D}_{\rm BS}^H\mathbf{D}_{\rm BS} \right)^{-1}\mathbf{D}_{\rm BS}^H \mathbf{H}^H\mathbf{D}_{\rm MS}\mathbf{D}_{\rm MS}^H
\mathbf{H}^H\mathbf{D}_{\rm MS}\right] \; .
\end{array}
\label{eq:ASE_UL}
\end{equation}
with $P_{\rm T, MS}$ the transmitted power at the MS and $P_{\rm T, BS}$ the transmitted power at the BS.
In the multiuser scenario the spectral efficiency on the uplink and downlink are respectively
\begin{equation}
\begin{array}{lll}
\!\!\mathcal{R}_{\rm DL}\!=\!\!\! \ds \sum_{k=1}^{K} \! \log_2 \det \!\!\left[ \mathbf{I}_M \!+\! \mathbf{R}_{k, {\rm DL}}^{-1}\mathbf{D}_{k}^H\mathbf{H}_{k}\mathbf{F}_{k}\mathbf{P}_{{\rm BS},k}\mathbf{F}_{k}^H\mathbf{H}_{k}^H\mathbf{D}_{k}\right] \; ,
\end{array}
\label{eq:ASE_DL_multiuser}
\end{equation}
and 
\begin{equation}
\begin{array}{lll}
\!\!\mathcal{R}_{\rm UL}\!= \!\!\!\ds \sum_{k=1}^{K}  \log_2 \det\!\! \left[ \mathbf{I}_M\! +\! \mathbf{R}_{k, {\rm UL}}^{-1}\mathbf{J}_{k}^H\mathbf{H}_{k}^H\mathbf{D}_{k}\mathbf{P}_{{\rm MS},k}\mathbf{D}_{k}^H\mathbf{H}_{k}\mathbf{J}_{k}\right] \; ,
\end{array}
\label{eq:ASE_UL_multiuser}
\end{equation}
where  $\mathbf{J}_{k}= \left[\widehat{\lambda_{k,1}  \mathbf{v}}_{k,1}, \ldots \widehat{\lambda_{k,M}  \mathbf{v}}_{k,M}\right]$ denotes the matrix containing the estimates of the products $\lambda_{k,i} \mathbf{v}_{k,i}$, 
$\mathbf{P}_{{\rm BS},k}$ is the  $(M \times M)$-dimensional diagonal matrix containing the power transmitted from the BS to the $k$-th MS for each stream, i.e. $\left[\mathbf{P}_{{\rm BS},k}\right]_{j,j}=\ds \frac{P_{\rm T, BS}}{M K \|\left[\mathbf{J}_{k}\right]_{(:,j)}\|^2}$, $\mathbf{P}_{{\rm MS},k}$ is the square $M$-dimensional diagonal matrix containing the power transmitted from the $k$-th MS to the BS for each stream, i.e. $\left[\mathbf{P}_{{\rm MS},k}\right]_{j,j}= \ds \frac{P_{\rm T, MS}}{M \|\left[\mathbf{D}_{k}\right]_{(:,j)}\|^2}$, and the matrices $\mathbf{R}_{k, {\rm DL}}$ and $\mathbf{R}_{k, {\rm UL}}$ are expressed as
\begin{equation}
\mathbf{R}_{k, {\rm DL}}= \sum_{\ell \neq k} \mathbf{D}_{k}^H\mathbf{H}_{k}\mathbf{J}_{\ell}\mathbf{P}_{{\rm BS},\ell} \mathbf{J}_{\ell}^H \mathbf{H}_{k}^H \mathbf{D}_{k},
\end{equation}
and
\begin{equation}
\mathbf{R}_{k, {\rm UL}}= \sum_{\ell \neq k} \mathbf{J}_{k}^H\mathbf{H}_{\ell}^H\mathbf{D}_{\ell}\mathbf{P}_{{\rm MS},\ell} \mathbf{D}_{\ell}^H \mathbf{H}_{\ell} \mathbf{J}_{k}.
\end{equation}

We will also represent, finally, the symbol error probability assuming differential phase shift keying signaling and differential non-coherent detection; indeed, notice that the proposed channel estimation methods do not rely on known training symbols, and the dominant eigenvectors are estimated with no information on the signal phase.

\subsection{Numerical results}
In our simulation setup, we consider a communication bandwidth of $W = 500$ MHz centered
over the carrier frequency $f_0=73$ GHz. The distance between the transmitter and the receiver is 50 m; the additive thermal noise is assumed to have a power spectral density of $-174$ dBm/Hz, while the front-end receiver at the BS and at the MS is assumed to have a noise figure of $6$ dB. The shown results come from an average over 500 random scenario realizations with independent channels. The remaining system parameters are reported in the figures' captions.

\begin{figure*}[t]
\centering
\includegraphics[scale=0.6]{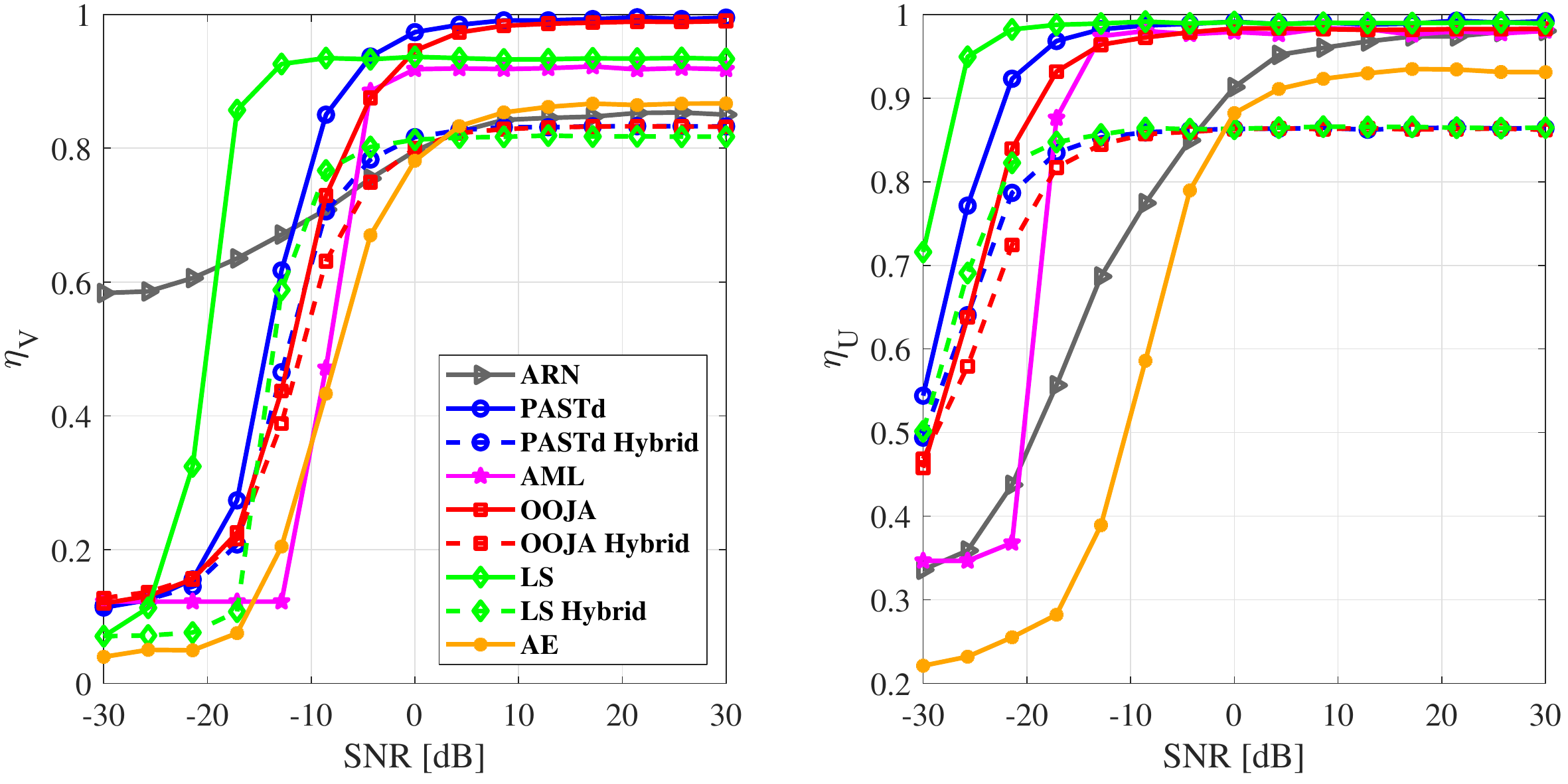}
\caption{$\eta_U$ and $\eta_V$ versus signal to noise ratio for a system with $N_{\rm MS} \times N_{\rm BS}=16 \times 64$. Parameters: $M=1$, $P_{\rm BS}=P_{\rm MS}=30$. For the HY implementations we have used $N_{\rm MS}^{\rm RF}=8$ and  $N_{\rm BS}^{\rm RF}=8$.}
\label{Fig:Correlation_SNR}
\end{figure*}

\begin{figure*}[t]
\centering
\includegraphics[scale=0.6]{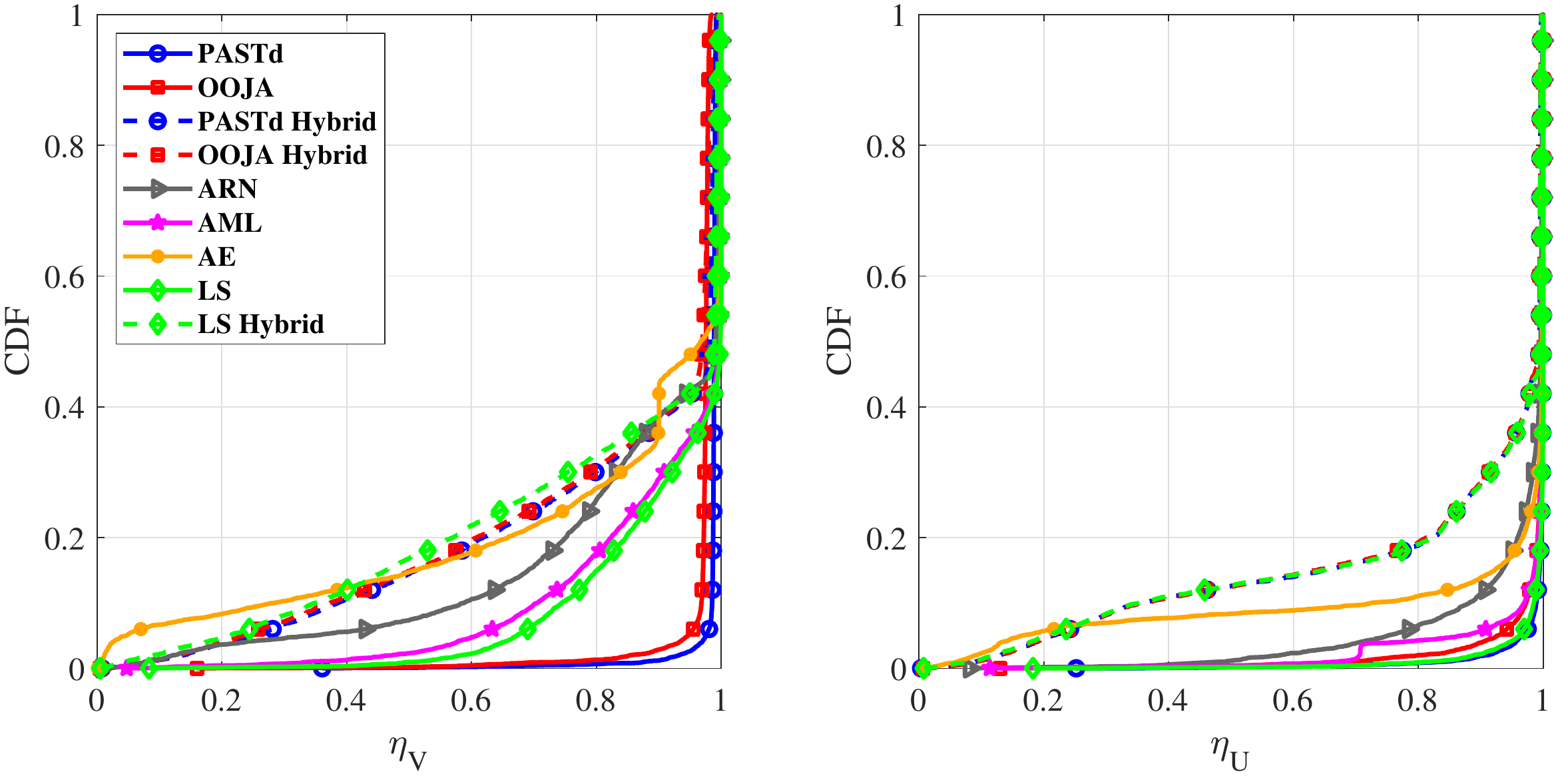}
\caption{CDF of $\eta_U$ and $\eta_V$ for a system with $N_{\rm MS} \times N_{\rm BS}=16 \times 64$. Parameters: $M=1$, $P_{\rm BS}=P_{\rm MS}=30$, the signal to noise ratio is 3 dB. For the HY implementations we have used $N_{\rm MS}^{\rm RF}=8$ and  $N_{\rm BS}^{\rm RF}=8$.}
\label{Fig:Correlation_CDF}
\end{figure*}

\begin{figure*}[t]
\centering
\includegraphics[scale=0.6]{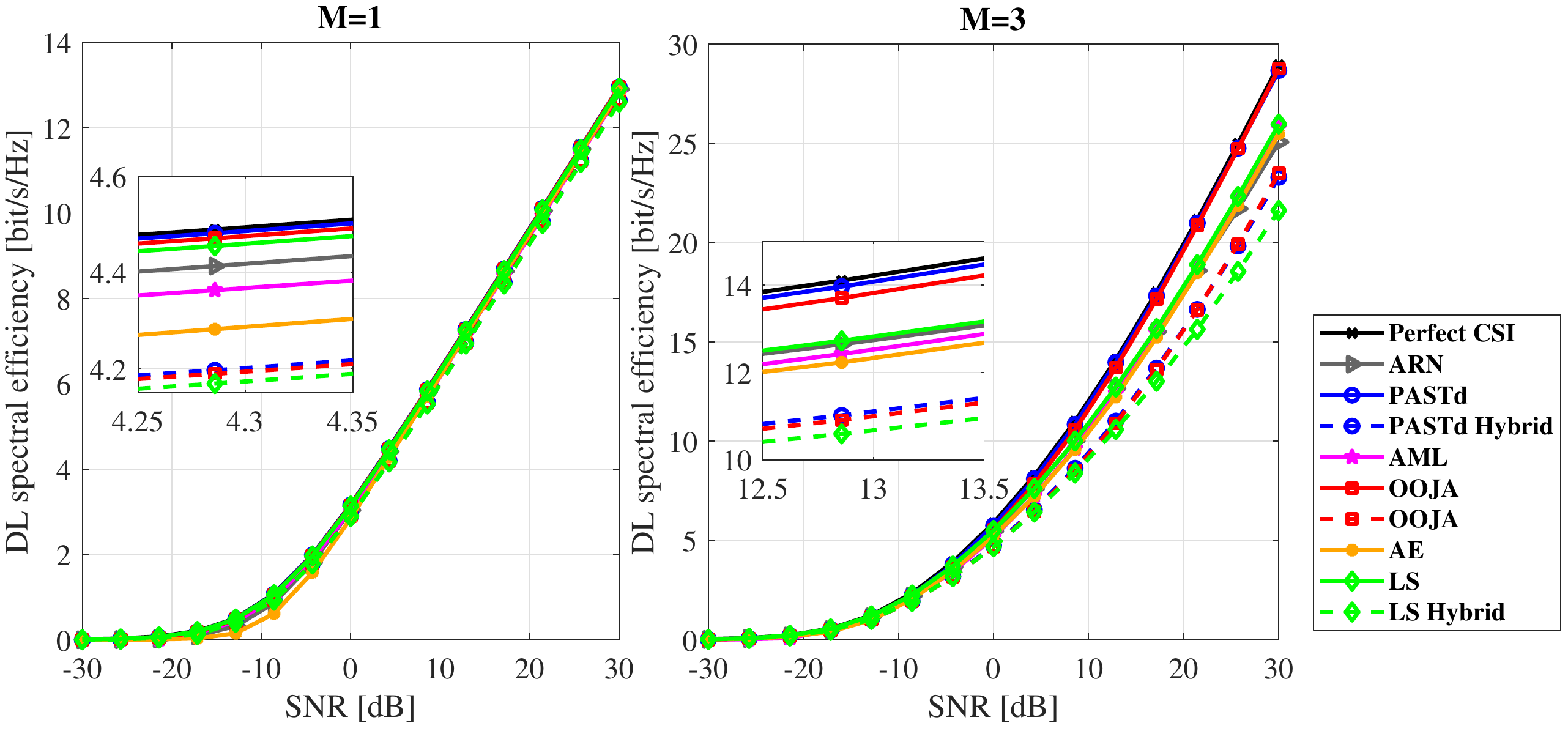}
\caption{Spectral efficiency versus signal to noise ratio for a system with $N_{\rm MS} \times N_{\rm BS}=16 \times 64$ and $P_{\rm BS}=P_{\rm MS}=30$. For the HY implementations we have used $N_{\rm MS}^{\rm RF}=8$ and  $N_{\rm BS}^{\rm RF}=8$.}
\label{Fig:SE_SNR}
\end{figure*}

We start considering the single user scenario, in this case the length of the training phase for all the algorithms is $P_{\rm BS}=P_{\rm MS}=30$; of these training samples, the first ten are used to perform an SVD of the sample covariance matrix of the data, and the corresponding dominant eigenvector is used to initialize the PASTd and OOJA algorithms.
First of all we focus on the performance measures $\eta_U$ and $\eta_V$. 
In Figure \ref{Fig:Correlation_SNR} we report  the average $\eta_U$ and $\eta_V$ versus the received SNR for all the considered algorithms proposed in the paper and for competing alternatives, and in particular the AML, ARN and AE methods, whose details are provided  in the Appendix. The system parameters are fully specified in the figure's caption. Inspecting the curves, it is seen that the proposed algorithms achieve satisfactory performance. Indeed, for SNR larger than 0 dB the average values of $\eta_V$ and $\eta_U$ approach very closely the limiting value 1 for both the PASTd and the OOJA algorithms with FD beamforming. The LS-based algorithm, on the other hand, achieves its ceil for SNR larger than -15 dB; its limiting performance value, however, for $\eta_V$ is slightly smaller than one, and this may be due to the quantization error induced by too small values for $L_{\rm MS}$ and $L_{\rm BS}$. 
This behavior thus suggests that the LS-based procedure can be used with improved performance for low values of SNR, while, instead, for larger values of SNR it is convenient to use the subspace-based techniques.
When considering the HY beamformers, we see that the limiting performance values are in the range $[0.8, 0.9]$, i.e. they do not approach unity. This is due to the fact that the fixed analog beamformer introduces a constraint that prevents the convergence of the estimated singular vectors to the true ones. Figure \ref{Fig:SE_SNR}, to be commented later, will show that the impact of the constraint posed by the analog beamforming stage on the system SE is however rather limited. Finally, the results show that the competing alternatives have a performance inferior to that of the proposed approaches, and, among the alternatives, the AML algorithm achieves the best performance. 

While Figure \ref{Fig:Correlation_SNR} has shown the average values of $\eta_V$ and $\eta_U$ versus the SNR, 
in Figure \ref{Fig:Correlation_CDF}, we set the SNR at 3 dB, and report  the cumulative distribution function (CDF)  of $\eta_U$ and $\eta_V$  across the 500 independent trials. This figure helps understanding the algorithms' performance in extreme situations, since its permits visualizing the percentiles of the plotted performance measures. 
Results show that the subspace-based algorithms with FD beamforming are very effective and are able to provide good performance also in the presence of heavily degraded channels. Indeed, focusing on the 5$\%$-percentile, the PASTs and OOJA algorithms with FD beamforming achieve correlation values quite close to 0.95, for both the left and right singular vectors. The other algorithms, instead achieve inferior performance, and, thus, they are less resistant to adverse situations with very bad channel realizations. Another general trend that can be observed is that in general $\eta_U$ is larger than $\eta_V$ i.e. the proposed algorithms estimate with higher reliability the left singular vectors of the channel. This behavior can be explained by noticing that the dominant left singular vectors are estimated based on the transmission of $N_{\rm BS}$-dimensional pilots (see 
 \eqref{eq:receivedMS}), while, in the second phase, the right singular vectors of the channel are estimated at the BS based on the transmission of $M$-dimensional pilots (see \eqref{eq:receiveBS}), with $M$, usually much smaller than $N_{\rm BS}$, the multiplexing order. Of course, the proposed algorithms might be modified and estimate the channel right singular vectors using $N_{\rm BS}$-dimensional pilots, but this would lead to increased algorithm complexity. 
\begin{figure*}[t]
\centering
\includegraphics[scale=0.6]{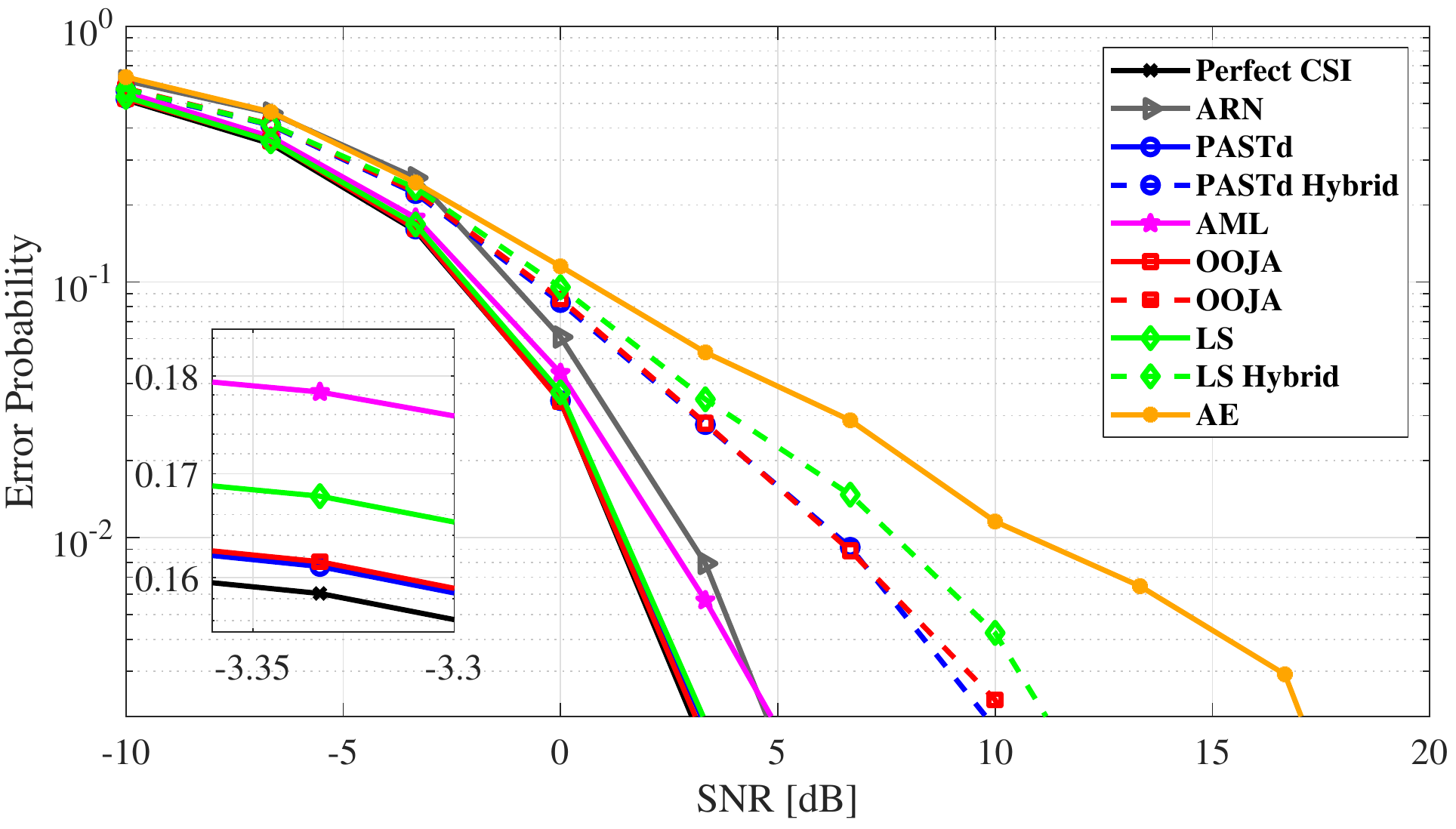}
\caption{Error probability for the BS-to-MS link versus received SNR. Parameters: 4-PSK differential modulation, $N_{\rm MS} \times N_{\rm BS}=16 \times 64$, $P_{\rm BS}=P_{\rm MS}=30$ and $M=1$. For the HY implementations we have used $N_{\rm MS}^{\rm RF}=8$ and  $N_{\rm BS}^{\rm RF}=8$.}
\label{Fig:Err_prob}
\end{figure*}

So far, the parameters $\eta_V$ and $\eta_U$ have been used to assess the performance of the proposed channel estimation algorithms. It is however of great interest to understand what is the impact of the proposed algorithms on one of the crucial performance measures for any wireless link, i.e. the achievable spectral efficiency. 
Figure \ref{Fig:SE_SNR} thus  reports the downlink achievable spectral efficiency evaluated as in \eqref{eq:ASE_DL}, for the BS-to-MS link versus SNR, for multiplexing order $M=1$ and $M=3$. 
For benchmarking purposes, we also report the curve corresponding to the case of perfect channel knowledge.
The results again confirm that the subspace-based techniques achieve the best performance, and are capable of attaining spectral efficiency values very close to those attainable in the ideal case of perfectly known channel. While for multiplexing order $M=1$ the several algorithms achieve very similar performance, for $M=3$ there is instead some evident performance gap among them. As an instance, for SNR=13 dB, the PASTd algorithm with FD beamforming achieves an average spectral efficiency of 14 bit/s/Hz, while the same algorithm with HY beamforming the average spectral efficiency is 11 bit/s/Hz, with a $-20\%$ loss. This gap, however, can be reduced by increasing the number of RF chains at the BS and at the MS. 

Figure \ref{Fig:Err_prob} reports the symbol error probability assuming differential uncoded  4-PSK modulation versus SNR for the BS-to-MS link and assuming multiplexing order $M=1$. Results show that the proposed algorithms with FD beamforming are fractions-of-dB far from the ideal curce corresponding to the case in which the channel is perfectly known. The use of HY beamformers, instead, introduces some performance degradation, that, at $10^{-2}$ error probability, can be quantified in approximately 5 dB. 

\begin{figure*}[t]
\centering
\includegraphics[scale=0.6]{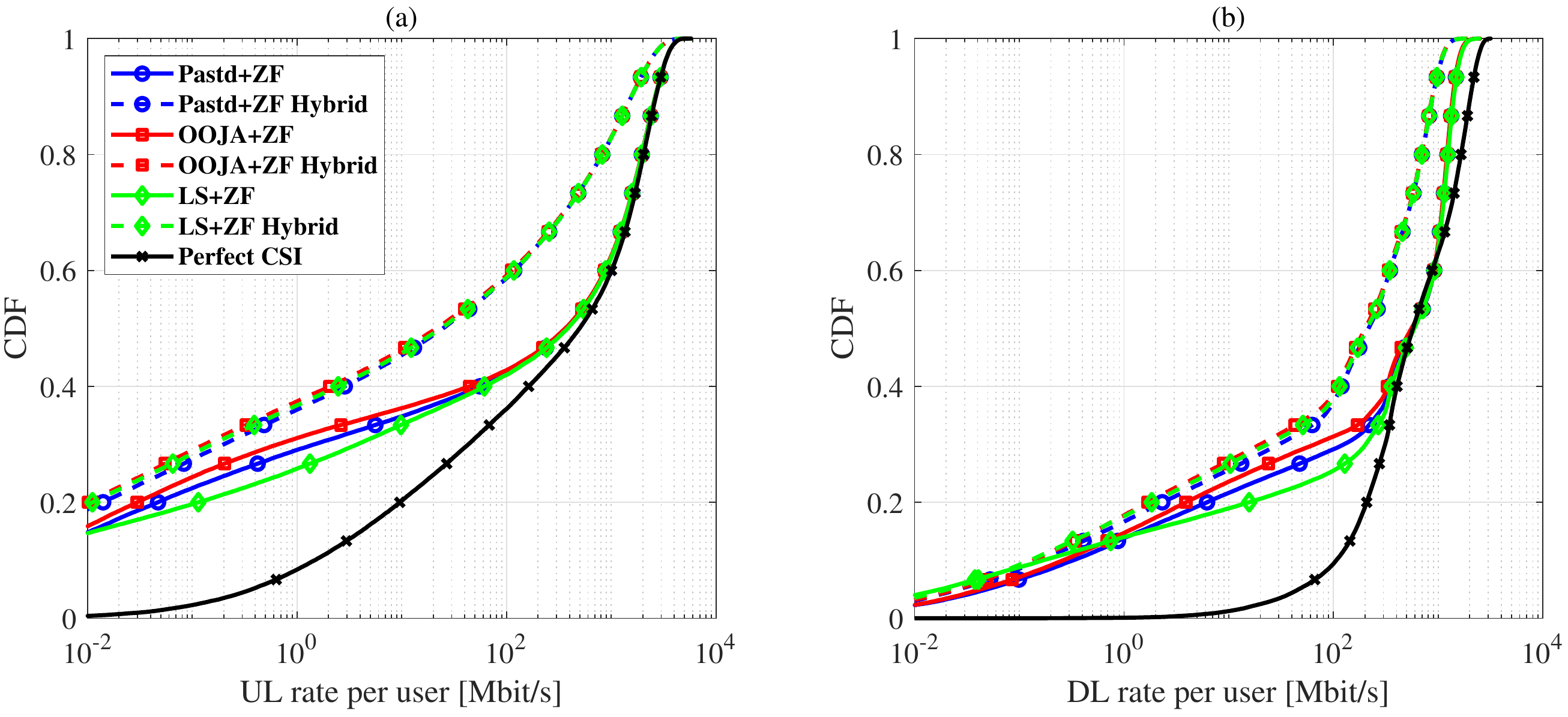}
\caption{CDF of the downlink (subfigure $(a)$) and uplink (subfigure $(b)$) achievable rate-per-user for a system with $K=15$ users with ZF approach for the dominant right channel eigenvectors estimation. We considered a system with $N_{\rm BS}=64$ and $N_{MS}=4$; for the HY-BF case we have used $N_{\rm BS}^{\rm RF}=16$ and  $N_{\rm MS}^{\rm RF}=2$, $P_{\rm BS}=60$, $P_{\rm MS}=32$, $P_{\rm T, MS}=-10$ dBW and $P_{\rm T, BS}=0$ dBW . The multiplexing order is $M=1$.}
\label{Fig:DL_UL_Multiuser}
\end{figure*}

Finally, we focus on the multiuser scenario and consider a system with one BS and $K=15$ MSs. The distance between the BS and the MSs was randomly chosen in the range $[5, 100]$ m. During phase (a) the BS transmit power is 1 W and the pilot length is $P_{\rm BS}=60$; in phase (b), the MSs transmit with a power of 0.1 W, and the pilot length is $P_{\rm MS}=32$. During the data communication phase the BS and MSs transmit powers are again 1 W and 0.1 W, respectively.
 Figure \ref{Fig:DL_UL_Multiuser}(a) shows the CDF of the downlink achievable rate-per-user of the system, while Figure \ref{Fig:DL_UL_Multiuser}(b) shows the CDF of the uplink achievable rate-per-user considering the algorithms proposed in Section V. Also in this case   we assume that the first ten training samples are used to perform an SVD of the sample covariance matrix of the data, and the corresponding dominant eigenvector is used to initialize the PASTd and OOJA algorithms at the MS. Results confirm that the proposed subspace-tracking algorithms exhibit very good performance, especially in their FD implementation. 
Results show that with just 4 antennas and 2 RF chains at the MSs it is possible to have, with the HY beamformers, data-rates in the range 100 Mbit/s - 1 Gbit/s for the 60 $\%$ and 40 $\%$ of the users in the downlink and in the uplink, respectively. Moreover, if we focus on the median rate, it can be seen that on the downlink the proposed algorithms with FD beamforming perform very similarly to the ideal system that assumes perfect channel knowledge, with a median rate equal to 600 Mbit/s,  while, instead, with HY beamforming this rate gets reduces to approximately 200 Mbit/s. On the uplink, instead, the median rate penalty caused by the use of HY beamforming is about one order of magnitude, and this loss can be mitigated by increasing the number of RF chains at the MSs.

\section{Conclusion}
This paper has been focused on the problem of channel estimation for wireless MIMO links at mmWave frequencies. Exploiting the clustered propagation channel model, it is shown that subspace tracking algorithms and the LS approach can be used in order to estimate the principal eigenvectors of the channel matrix. The proposed algorithms have been extended to the case in which HY analog/digital beamforming is used, and to a single-cell multiuser scenario. Results have shown that the proposed estimation algorithms, and in particular the ones based on the subspace tracking algorithms, are effective and capable of attaining good performance levels with low complexity. 
Results have also shown that there is a noticeable performance gap of HY beamformers with respect to the case of FD beamforming, and this suggests that the design of optimized low-complexity beamformers is one of the research areas that are worth being investigated, at least until the technology will not permit the use of FD beamformers. 
Further research in this area is also focused on the extension of the proposed algorithms to the case in which 3D-lens antennas arrays and parasitic arrays are employed, as well as to their application in a cell-free massive MIMO settting \cite{Alonzo_Buzzi_TGCN}. 

\section{Acknowledgement}
The authors wish to thank the authors of papers \cite{haghighatshoar2016massive} and \cite{ghauch2016subspace} for providing helpful assistance on the Matlab programming of their proposed channel estimation algorithms. 

This work has been supported by the MIUR program ``Dipartimenti di Eccellenza 2018-2022".

\section*{Appendix} \label{Appendix_Algorithms}
In this Appendix, we provide the details about the 
generalization of the AML algorithm in \cite{haghighatshoar2016massive} to the case in which
multiple antennas are used at both sides of the communication link. Then, we provide very few details on two other channel estimation procedures that we have used for benchmarking purposes. 

\subsection{AML Algorithm}
For the multiantenna case, the AML algorithm can be described as follows. 
In the first part of the training interval the BS transmits its pilot signals for $n=1, \ldots, P_{\rm BS}$. Denoting by $\bar{\mathbf{s}}_{\rm BS}(n)$ the $(N_{\rm BS} \times 1)$-dimensional vector transmitted by the BS at the discrete epoch $n$, the received discrete-time signal at the mobile station is represented by the following $(N_{\rm MS} \times 1)$-dimensional vector:
\begin{equation}
\bar{\mathbf{r}}_{\rm MS}(n) = \mathbf{H}\bar{\mathbf{s}}_{\rm BS}(n) + \mathbf{w}_{\rm MS}(n),
\end{equation}  
where $\mathbf{H}$ is the $(N_{\rm MS} \times N_{\rm BS})$-dimensional channel matrix with the structure reported in \eqref{eq:channel1} and $\mathbf{w}_{\rm MS}(n)$ is the $(N_{\rm MS} \times 1)$-dimensional noise vector and $\bar{\mathbf{s}}_{\rm BS}(n)$ is the $(N_{\rm BS} \times 1)$-dimensional pilot signal, whose entries are all equal to one.
Let $\mathbf{T}_{\rm MS}=\mathbf{H}\mathbf{H}^H=\widetilde{\mathbf{U}}\widetilde{\mathbf{\Lambda}}^2 \widetilde{\mathbf{U}}^H$, denote by $\mathbf{B}_{\rm MS }$ the $(N_{\rm MS}^{\rm RF} \times N_{\rm MS})$-dimensional 0/1 sampling operator, and consider the  $N_{\rm MS}^{\rm RF}$-dimensional projection of $\bar{\mathbf{r}}_{\rm MS}(n)$ onto $\mathbf{B}_{\rm MS}$ as:
\begin{equation}
\bar{\mathbf{x}}_{\rm MS}(n) = \mathbf{B}_{\rm MS }\bar{\mathbf{r}}_{\rm MS}(n)
\end{equation}
Let $\mathbf{C}_{x, \rm{MS}}=\frac{1}{P_{\rm BS}}\sum_{n=1}^{P_{\rm BS}}{\bar{\mathbf{x}}_{\rm MS}(n)\bar{\mathbf{x}}_{\rm MS}^H(n)}$ be the covariance matrix of the sketches $\bar{\mathbf{x}}_{\rm MS}(n)$, let $\mathbf{C}_{x, \rm{MS}}=\mathbf{P}\mathbf{L}\mathbf{P}^H$ be its SVD, and define $\mathbf{\Delta}_{\rm MS}=\mathbf{P}\mathbf{L}^{1/2}$. The AML algorithm proposed in \cite{haghighatshoar2016massive} is cast as the following semi-definite program (SDP)
\begin{equation}
\begin{array}{llll}
 (\mathbf{S}_{\rm MS}^*,\mathbf{K}_{\rm MS}^*)= & \!\!\!\!\!\!\!\!\!\!\!\underset{\mathbf{M} \in \mathbb{T}_+,\mathbf{K}_{\rm MS} \in \mathbb{C}^{N_{\rm MS}^{\rm RF} \times N_{\rm MS}^{\rm RF}}}{\arg\min} \!\!\!\!\!\!\!\!\!\text{tr}\left(\mathbf{B}_{\rm MS }\mathbf{M}\mathbf{B}_{\rm MS }^H\right) + \text{tr} \left(\mathbf{K}_{\rm MS}\right) \\
\text{subject to}  &
 \left[ \begin{matrix}
 \sigma^2_n\mathbf{I}_{N_{\rm MS}^{\rm RF}}+\mathbf{B}_{\rm MS }\mathbf{M}\mathbf{B}_{\rm MS }^H & & & \mathbf{\Delta}_{\rm MS} \\
 \mathbf{\Delta}_{\rm MS}^H & & & \mathbf{K}_{\rm MS}
 \end{matrix} \right] \succeq 0,
\end{array}
\label{AML_SDP}
\end{equation} 
where $\mathbb{T}_+$ denotes the space of all $N_{\rm MS} \times N_{\rm MS}$ Hermitian PSD Toeplitz matrices, and $\sigma^2_n$ is an estimate of the noise variance in each antenna element. The optimal solution of \eqref{AML_SDP} gives an estimate of the matrix $\mathbf{T}_{\rm MS}$. Since solving the SDP \eqref{AML_SDP} is a time-consuming task, especially for a large number of antennas, in \cite{haghighatshoar2016low} the authors provide a low-complexity algorithm to solve \eqref{AML_SDP}. Let $\mathcal{G}_{\rm MS}$ a discrete grid of size $G_{\rm MS}$ over the angular range $\left[-\frac{\pi}{2}, \frac{\pi}{2}\right]$. We assume that $\mathcal{G}_{\rm MS}$ consists of uniformly spaced angles $\theta_i^{\rm MS}=\left(-1+\frac{2(i-1)}{G_{\rm MS}}\right)\frac{\pi}{2}$, for $i=1, \ldots, G_{\rm MS}$. Denote by $\bar{\mathbf{G}}_{\rm MS}$  the $(N_{\rm MS} \times G_{\rm MS})$-dimensional matrix whose columns are given by $\mathbf{a}_{\rm MS}(\theta_i^{\rm MS})$, corresponding to the array response at the MS at the AoA $\theta_i^{\rm MS} \in \mathcal{G}_{\rm MS}$. In \cite{haghighatshoar2016low} the authors assume that $\mathcal{G}_{\rm MS}$ is dense enough such that every matrix $\mathbf{T}_{\rm MS}$ can be well approximated by
\begin{equation}
\begin{array}{llll}
\mathbf{T}_{\rm MS} &\approx \bar{\mathbf{G}}_{\rm MS} \text{diag}\left(t_1, \ldots, t_{G_{\rm MS}}\right) \bar{\mathbf{G}}_{\rm MS}^H \\&= \sum_{i=1}^{G_{\rm MS}}{t_{i}\mathbf{a}_{\rm MS}\left(\theta_i^{\rm MS}\right)\mathbf{a}_{\rm MS}^H(\theta_i^{\rm MS})},
\end{array}
\label{S_approx}
\end{equation} 
with appropriate $t_{i} \geq 0, i=1,\ldots, G_{\rm MS}$.  Now, consider the following convex optimization problem to be solved for the variable $\mathbf{W}_{\rm MS} $:
\begin{equation}
\!\!\mathbf{W}_{\rm MS}^*\!\!=\!\! \underset{\mathbf{W}_{\rm MS}}{\arg\min} \frac{1}{2}\|\widetilde{\mathbf{G}}_{\rm MS}\mathbf{W}_{\rm MS}\!-\!\mathbf{X}_{\rm MS}\|^2\!\!+\!\!\sigma^2_n\sqrt{P_{\rm BS}}\|\mathbf{W}_{\rm MS}\|_{2,1},
\label{conv_opt}
\end{equation}
where $\mathbf{W}_{\rm MS}$ is a $(G_{\rm MS} \times N_{MS}^p)$-dimensional matrix, with an $\ell_{2,1}$ norm defined by $\|\mathbf{W}_{\rm MS}\|_{2,1}=\sum_{i=1}^{G_{\rm MS}}{\|\mathbf{W}_{\rm{MS},(i,.)}\|}$,  $\mathbf{X}_{\rm MS}=\left[\bar{\mathbf{x}}_{\rm MS}(1), \ldots, \bar{\mathbf{x}}_{\rm MS}(P_{\rm BS})\right]$ is the $(N_{\rm MS}^{\rm RF} \times P_{\rm BS})$-dimensional matrix of noisy sketches, and $\widetilde{\mathbf{G}}_{\rm MS}=\frac{1}{\sqrt{N_{\rm MS}^{\rm RF}}}\mathbf{B}_{\rm MS }\bar{\mathbf{G}}_{\rm MS}$. In \cite{haghighatshoar2016low}, the authors show that, if we suppose that the grid $\mathcal{G}_{\rm MS}$ is dense enough such that every matrix $\mathbf{T}_{\rm MS}$ can be precisely approximated according to \eqref{S_approx}, the SDP \eqref{AML_SDP} and the convex optimization \eqref{conv_opt} are equivalent. Then, the optimal solution of \eqref{AML_SDP} can be approximated by $\mathbf{T}_{\rm MS}^* = \bar{\mathbf{G}}_{\rm MS} \text{diag}\left(t_1^*, \ldots, t_{G_{\rm MS}}^*\right) \bar{\mathbf{G}}_{\rm MS}^H$, where $t^*_i=\frac{1}{\sqrt{N_{\rm MS}^{\rm RF}}}\|\mathbf{W}_{\rm{MS},(i,.)}^*\|$. 

The \textit{forward-backward splitting} (FBS) can be used for minimizing the sum of two convex functions \cite{combettes2011proximal}. Indeed, after suitable scaling, we can write the objective function \eqref{conv_opt} as the sum of two convex function 
\begin{equation}
f\left(\mathbf{W}_{\rm MS}\right)=f_1\left(\mathbf{W}_{\rm MS}\right)+f_2\left(\mathbf{W}_{\rm MS}\right),
\end{equation}
where $f_1\left(\mathbf{W}_{\rm MS}\right)=\frac{1}{2\zeta}\|\tilde{\mathbf{G}}_{\rm MS}\mathbf{W}_{\rm MS}-\mathbf{X}_{\rm MS}\|^2$, $ f_2\left(\mathbf{W}_{\rm MS}\right)=\|\mathbf{W}_{\rm MS}\|_{2,1}$, and $\zeta=\sigma_n^2 \sqrt{N_{\rm MS}^p}$.

The gradient of $f_1$ is given by 
\begin{equation}
\nabla f_1\left(\mathbf{W}_{\rm MS}\right)=\frac{1}{\zeta}\widetilde{\mathbf{G}}_{\rm MS}^H\left(\tilde{\mathbf{G}}_{\rm MS}\mathbf{W}_{\rm MS}-\mathbf{X}_{\rm MS}\right).
\label{grad_f1}
\end{equation}
The function $\nabla f_1(\cdot)$ defined in  \eqref{grad_f1} is a Lipschitz function\footnote{A real-valued function $f(\cdot)$ is said to be a 
Lipschitz function if $|f(x)-f(y)|\leq C |x-y|$, for any value of $x$ and $y$, and with $C$ a constant independent of 
$x$ and $y$.} 
with  Lipschitz constant $\kappa$, i.e,
\begin{equation}
\|\nabla f_1\left(\mathbf{W}_{\rm MS}\right)-\nabla f_1\left(\mathbf{W}_{\rm MS}'\right)\| \leq \kappa \|\mathbf{W}_{\rm MS}-\mathbf{W}_{\rm MS}'\|,
\end{equation}
with $\kappa=\frac{1}{\zeta}\lambda_{\text{max}}\left(\widetilde{\mathbf{G}}_{\rm MS}^H\widetilde{\mathbf{G}}_{\rm MS}\right)=\frac{1}{\zeta}\lambda_{\text{max}}\left(\widetilde{\mathbf{G}}_{\rm MS}\widetilde{\mathbf{G}}_{\rm MS}^H\right)$, and $\lambda_{\text{max}}(\cdot)$ denoting the maximum singular value function. 
The AML procedure with FBS for the channel subspace estimation at the MS is reported in Algorithm \ref{AML_MS}, where we have:
\begin{equation}
\text{prox}_{\alpha,f_2}\left(\mathbf{W}_{\rm MS}\right)_{i,.}=\frac{\left(\|\mathbf{W}_{\rm{MS},(i,.)}\|-\alpha\right)_+}{\|\mathbf{W}_{\rm{MS},(i,.)}\|}\|\mathbf{W}_{\rm{MS},(i,.)}\|,
\end{equation}
with $(x)_+ = \max{(x,0)}$.
\begin{algorithm}[!t]

\caption{The AML Algorithm with FBS for the estimation at the MS}

\begin{algorithmic}[1]

\label{AML_MS}

\STATE Fix $\mathbf{W}_{\rm MS}^{(0)}$, set $\mathbf{Z}_{\rm MS}^{(0)}=\mathbf{W}_{\rm MS}^{(0)}$, and $t_0=1$

\FOR { $k=0,1,\ldots,$}

\STATE  {$\mathbf{R}_{\rm MS}^{(k)}=\mathbf{Z}_{\rm MS}^{(k)}-\frac{1}{\beta}\nabla f_1\left(\mathbf{Z}_{\rm MS}^{(k)}\right)$}

\STATE {$\mathbf{W}_{\rm MS}^{(k+1)}=\text{prox}_{\frac{1}{\beta},f_2}\left(\mathbf{R}_{\rm MS}^{(k)}\right)$}

\STATE {$t_{k+1}=\frac{1+\sqrt{4t_k^2+1}}{2}$}

\STATE {$\alpha_k=1+\frac{t_k-1}{t_{k+1}}$}

\STATE {$\mathbf{Z}_{\rm MS}^{(k+1)}=\mathbf{W}_{\rm MS}^{(k)}+\alpha_k\left(\mathbf{W}_{\rm MS}^{(k+1)}-\mathbf{W}_{\rm MS}^{(k)}\right)$}

\ENDFOR

\end{algorithmic}

\end{algorithm}

Following similar steps, that we do not report for the sake of brevity, it is possible to perform channel subspace estimation at the MS also.

\subsection{ARN Algorithm}
The ARN algorithm for channel subspace estimation has been proposed in  \cite{ghauch2016subspace}. The method is  based on the Arnoldi iteration, exploiting channel reciprocity in TDD systems and the sparsity of the channel's eigenmodes. The BS selects a random unit beamforming vector and sends a pilot signal to the MS. The signal received by the MS is echoed back to the BS, in an Amplify-and-Forward like fashion. Then, exploiting channel reciprocity, the received signal at the BS is first normalized and then echoed back to the MS. This procedure is done iteratively several times, and leads to the estimation of the matrix $\mathbf{D}_{\rm BS}$. A similar procedure can be applied using MS-initiated echoing, to obtain $\mathbf{D}_{\rm MS}$ at the MS. We refer to the paper \cite{ghauch2016subspace} for the full details.

%This procedure is done iteratively. With this procedure, we are able to estimate $\mathbf{Q}_{\rm BS}$. The SE-ARN procedure used in this paper is reported in Algorithm \ref{SE_ARN}. This same procedure can be applied using MS-initiated echoing, to obtain $\mathbf{D}_{\rm MS}$ at the MS.
%\begin{algorithm}[!t]
%
%\caption{The SE-ARN for the estimation at the BS}
%
%\begin{algorithmic}[1]
%
%\label{SE_ARN}
%
%\STATE Set random unit norm vector $\mathbf{b_1}; \; \; \mathbf{B}=\left[\mathbf{b}_1\right]$
%
%\FOR { $\ell=1,\ldots, N_{\rm MS}^p$}
%
%\STATE  {$\mathbf{s}_{\ell}=\mathbf{H}\mathbf{b}_{\ell}+\mathbf{w}_{\ell,\rm {MS}}$}
%
%\STATE {$\mathbf{p}_{\ell}=\mathbf{H}^H\mathbf{s}_{\ell}+\mathbf{w}_{\ell,\rm {BS}}$}
%
%\STATE {$t_{k,\ell}=\mathbf{b}_{\ell}\mathbf{p}_{\ell} \; \; \forall k=1,\ldots, \ell$}
%
%\STATE {$\mathbf{r}_{\ell}=\mathbf{p}_{\ell}-\sum_{k=1}^{\ell}{\mathbf{b}_{k}t_{k,\ell}}$}
%
%\STATE {$t_{\ell+1,\ell}=\|\mathbf{r}_{\ell}\|_2$}
%
%\STATE {$\mathbf{B}=\left[\mathbf{B}, \mathbf{b}_{\ell+1}=\frac{\mathbf{r}_{\ell}}{t_{\ell+1,\ell}}\right]$}
%\ENDFOR
%\STATE {$\mathbf{T}=\tilde{\mathbf{\Theta}}\tilde{\mathbf{\Lambda}}\tilde{\mathbf{\Theta}}^{-1}$}
%\STATE {$\mathbf{Q}_{\rm MS}=\text{qr}(\mathbf{B}\tilde{\mathbf{\Theta}}_{1:M})$}
%\end{algorithmic}
%
%\end{algorithm}

\subsection{AE algorithm}
The Adaptive Estimation (AE) algorithm  was introduced in \cite{alkhateeb2014channel}. The paper first formulates and develops a hierarchical multiresolution codebook based on HY analog/digital precoding, and then proposes the AE algorithm using the codebook previously determined. In our numerical simulations,  we use the simulation code available at \cite{URL_Code_heath}; we refer to paper  \cite{alkhateeb2014channel} for the full details on the AE algorithm.

\bibliographystyle{IEEEtran}
%\nocite{*}

%\linespread{1.33}
\bibliography{FracProg_SB,finalRefs,references}

\end{document}